\documentclass[fleqn,usenatbib]{mnras}

\usepackage{newtxtext,newtxmath}
\usepackage[T1]{fontenc}
\usepackage{ae,aecompl}
\usepackage{times}

\usepackage{graphicx}	
\usepackage{amsmath}	
\usepackage{amssymb}	
\usepackage{ulem}
\usepackage[ampersand]{easylist}

\usepackage{latexsym}
\usepackage{enumerate}
\usepackage{amstext}
\usepackage{morefloats}

\usepackage{natbib}
\usepackage{rotating}
\usepackage{changepage}
\usepackage{scrextend}
\usepackage[ampersand]{easylist}
\usepackage{xcolor}
\usepackage{ulem}
\usepackage{bigints}

\newcommand{\bt}[1]{{\textcolor{blue}{{#1}}}}

\title[Magnesium and europium in IllustrisTNG]{First results from the IllustrisTNG simulations: A tale of two elements -- chemical evolution of magnesium and europium}

\author[J.~P.~Naiman et al.]{Jill~P.~Naiman$^{1}$\thanks{E-mail: jill.naiman@cfa.harvard.edu}, 
Annalisa Pillepich$^{2}$, Volker Springel$^{3,4}$, Enrico Ramirez-Ruiz$^{5}$, \newauthor Paul Torrey$^{6}$, Mark Vogelsberger$^{6}$, R\"udiger~Pakmor$^3$, Dylan Nelson$^{7}$, 
\newauthor Federico Marinacci$^{6}$, Lars Hernquist$^{1}$, Rainer Weinberger$^{3}$, Shy Genel$^{8}$
\vspace*{0.2cm} \\
$^{1}$Harvard-Smithsonian Center for Astrophysics, 60 Garden Street, Cambridge, MA, 02138, USA\\
$^2$Max-Planck-Institut f\"ur Astronomie, K\"onigstuhl 17, 69117 Heidelberg, Germany\\
$^3$Heidelberger Institut f\"{u}r Theoretische Studien,
  Schloss-Wolfsbrunnenweg 35, 69118 Heidelberg, Germany\\
  $^4$Zentrum f\"ur Astronomie der Universit\"at Heidelberg, ARI, M\"onchhofstrasse
12-14, 69120 Heidelberg, Germany\\
  $^5$Department of Astronomy and Astrophysics, University of California, Santa Cruz, CA 95064, USA\\
  $^6$Department of Physics, Kavli Institute for Astrophysics and Space Research, MIT, Cambridge, MA 02139, USA\\
  $^7$Max-Planck-Institut f\"ur Astrophysik, Karl-Schwarzschild-Str. 1, D-85748, Garching, Germany\\
  $^8$Center for Computational Astronomy, Simons Foundation, New York, USA
}

\date{11 July 2017}

\pubyear{2017}

\begin{document}
\label{firstpage}
\pagerange{\pageref{firstpage}--\pageref{lastpage}}
\maketitle

\begin{abstract}

The distribution of elements in galaxies provides a wealth of information about their production sites and their subsequent mixing into the interstellar medium.  Here we investigate the elemental distributions of stars in the IllustrisTNG simulations. We analyze the abundance ratios of magnesium and europium in Milky Way-like galaxies from the TNG100 simulation (stellar masses ${\log} (M_\star / {\rm M}_\odot) \sim 9.7 - 11.2$). Comparison of observed magnesium and europium for individual stars in the Milky Way with the stellar abundances in our more than $850$ Milky Way-like galaxies provides stringent constraints on our chemical evolutionary methods. Here, we use the magnesium to iron ratio as a proxy for the effects of our SNII and SNIa metal return prescription and as a comparison to a variety of galactic observations. The europium-to-iron ratio tracks the rare ejecta from neutron star -- neutron star mergers, the assumed primary site of europium production in our models, and is a sensitive probe of the effects of metal diffusion within the gas in our simulations. We find that europium abundances in Milky Way-like galaxies show no correlation with assembly history, present day galactic properties, and average galactic stellar population age. We reproduce the europium-to-iron spread at low metallicities observed in the Milky Way, and find it is sensitive to gas properties during redshifts $z \approx 2-4$. We show that while the overall normalization of [Eu/Fe] is susceptible to resolution and post-processing assumptions, the relatively large spread of [Eu/Fe] at low [Fe/H] when compared to that at high [Fe/H] is quite robust.

\end{abstract}



\section{Introduction}
 
Chemical evolutionary studies allow us to probe the global enrichment history of the Universe and individual star formation histories of galaxies, with each element class containing information about different aspects of the star formation and galactic enrichment cycle~\citep[e.g.][]{rana1991, matteucci2012}. In particular, the abundance of $\alpha$-elements such as magnesium and oxygen has been demonstrated to be a sensitive proxy for the initial phases of stellar evolution, which are dominated by heavy element production and return from core-collapse supernovae \citep[SNII; e.g.][]{tinsley1980, kobayashi2006, woosley2015}. In addition, iron peak elements like iron, cobalt, and chromium are created over longer timescales in Type Ia supernovae (SNIa), whereas s-process neutron capture elements (e.g.~strontium and barium) probe their proposed injection sites -- intermediate and low-mass asymptotic giant branch stars \citep[AGB; e.g.][]{tinsley1980,pagel1997,matteucci2001}. Similarly, r-process elements (e.g.\ europium) provide a window into the most violent events in our Universe -- neutron star mergers \citep[e.g.,][]{latt1977,fre1999} and core-collapse supernovae \citep[e.g.,][]{tak1994, woosley1994}.

Each process of element injection is affected by a variety of galactic, stellar, and interstellar medium properties such as the shape of the initial mass function (IMF), delay time distribution of SNIa~\citep[e.g.,][]{wiersma2009, walcher2016}, nucleosynthesis yields \citep[e.g.][]{tinsley1979,prantzos1994,thomas1999}, and environmental parameters which determine how enriched and pristine gas is mixed before forming stars with a specific yield pattern~\citep[e.g.,][]{torrey2012}. Detailed measurements of abundance patterns in individual stars are, however, possible only in the Milky Way and nearby galaxies \citep[e.g.][]{tolstoy2009}, while the studies of further galaxies require averaging spectra over the entire galaxy's stellar population, and in many cases, stacking spectra of similar galaxies to generate high enough signal to noise data for robust analysis \citep[e.g.][]{oconnell1976,peterson1976,pel1989}.  Thus, comparison of simulated chemical evolution in the Milky Way is often the most robust test of our theoretical understanding and therefore the target of our present comparison.

Chemical enrichment models like those in our new IllustrisTNG suite of cosmological simulations allow us to track the average abundance patterns in galaxy populations and individual galaxies simultaneously.  This paper complements the analysis of semi-analytical models~\citep[e.g.,][]{yates2013, hirschmann2016, delucia2017} and simulations of individual galaxies \citep{guedes2011, vandevoort2015, shen2015, vandevoort2017} by providing a statistical analysis of how abundance patterns are affected by galaxy formation histories based on hundreds of simulated galaxies. 
Semi-analytical models offer the opportunity to explore many different methods of chemical enrichment for minimal computational cost.  These studies are especially useful when tracking the evolution of elements like europium in which many different pathways of enrichment have been proposed~\citep{ces2006}.  While such methods are useful for exploring the average chemical evolutionary properties of a galaxy's stellar population, they can not reproduce the spreads in abundance ratios at fixed [Fe/H] that are widely observed in the Milky Way \citep[e.g.][]{kobayashi2006}.   Here, hydrodynamical simulations of star clusters, galaxies, and galaxy groups can be utilized to model the different chemical evolutionary paths for different stars within the larger galactic population allowing for more variance in enrichment histories~\citep[e.g.][]{borg2008,matteucci2014}.
 
In this paper, we focus on elements which are sensitive indicators of enrichment histories and gas mixing properties, with an emphasis on the magnesium-to-iron and europium-to-iron ratios. Studies of magnesium and iron probe ejecta properties of core-collapse and Type Ia supernovae, and span size scales from individual stars \citep[e.g.][]{mcwilliam1997} to stacked spectra from multiple galaxies \citep{graves2008,thomas2010,joh2012,worthey2013,conroy2014}.  In contrast, investigations of europium are generally relegated to local sources \citep[e.g.][]{sneden2008,ji2016}.

In simulations which can resolve galactic evolutionary processes, the magnesium to iron ratio is often used as a test of the robustness of applied SNII and SNIa injection yields and methods in concert with larger scale feedback prescriptions \citep{shen2015, vandevoort2015}. Within a cosmological context, NSNS mergers are important because they are rare, turning their abundance variations into an insightful indicator of how mixing is treated in the simulations. In particular, we investigate the europium distribution in Milky Way-sized galaxies within the TNG100 simulation, and examine what this distribution can tell us about the gas mixing processes during star formation in our calculations. How well magnesium and europium distributions compare to observed trends provides sensitive tests of our yield tables, injection processes, star formation histories, and diffusion of metals in the gas phase.

The present paper concentrates on chemical enrichment and is one in a series of five papers introducing the IllustrisTNG project, each focusing on a different scientific aspect of the simulations. In \bt{Marinacci et al. (2017)}, we analyze the magnetic field strengths in haloes and compute radio synchrotron maps for clusters of galaxies. In \bt{Springel et al. (2017)} we examine the clustering of matter and galaxies far into the non-linear regime, while in \bt{Nelson et al. (2017)}, we compare the color bimodality distribution of our simulated galaxies to data from the SDSS. Finally, in \cite{methodspaper} we focus on the stellar mass content of massive galaxy groups and clusters.

This paper begins with a brief introduction of the new and updated methods for treating the creation and advection of metals within the IllustrisTNG simulations in Section~\ref{section:methods}. A brief summary of some of the salient issues regarding comparisons of simulated and observed abundance ratios is given in Section~\ref{section:observations}, a description of $\alpha$- and r-process enrichment within Milky Way-sized galaxies is presented in Section~\ref{section:mwabundances}, and a discussion of the origins of our simulated abundance ratios is provided in Section~\ref{section:origins}. We end the paper with a discussion of how resolution affects our results and a summary of our conclusions in Sections~\ref{section:discussion} and \ref{section:conclusions}, respectively.

 \begin{figure}
\centering
\includegraphics[width=0.42\textwidth]{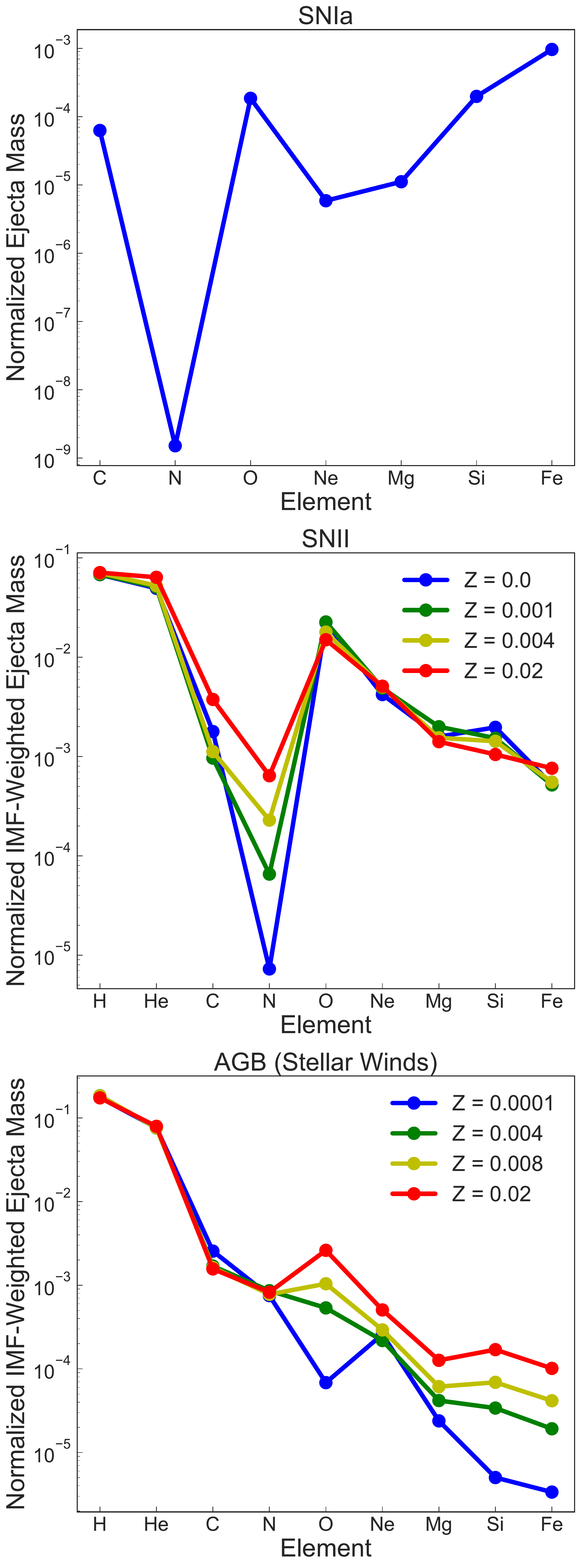}
\caption{The ejected mass per element for each process is shown as a percentage of the total mass of a single stellar population.  For SNIa (top panel) only 7 elements are shown as this process is assumed to eject no hydrogen or helium.  For the two bottom panels the ejecta for each element have been summed across the full mass range ($8-120 \, {\rm M}_\odot$ for SNII, $0.1-7.5 \, {\rm M}_\odot$ for AGB), after multiplication with the IMF used in the IllustrisTNG simulations \citep{chab2003}. }
\label{fig:yields}
\end{figure}

\section{Methodology} \label{section:methods}

\subsection{Chemical enrichment implementation in IllustrisTNG}

The methods used for production, deposition, and advection of elements in {\it The Next Generation Illustris} (IllustrisTNG)\footnote{\url{http://www.tng-project.org}} simulations have been updated from the original Illustris model described and tested in~\citet{vogelsberger2013} and~\citet{torrey2014}. A total of 9 individual elements are tracked explicitly (H, He, C, N, O, Ne, Mg, Si, Fe) plus a pseudo ``other metal'' species, which represents the sum of all other metals not individually tracked.  Throughout the simulation, mass and metals are returned to the interstellar medium (ISM) from aging stellar populations while accounting for heavy element nucleosynthesis and mass return in core-collapse supernovae, Type-Ia supernovae, and asymptotic giant branch (AGB) stars. 

As in Illustris \citep{vogelsberger2013}, the mass returned from the SNII and AGB channels for an element, $i$, during a time step, $\Delta t$, depends on the metallicity, $Z$ and age, $t$, of the single stellar population represented by each star particle:  
\begin{equation}
\Delta M_i(t, \Delta t, Z) = \bigintssss_{\mathcal{M}(t+\Delta t)}^{\mathcal{M}(t)} ( y_i(m,Z) + m \chi_i f_{\rm{rec}}(m,Z)) \Theta(m) dm 
\end{equation}
where the IMF, $\Theta(m)$, is taken to be that of \citet{chab2003}, $f_{\rm{rec}}$ is the fraction of mass returned to the ISM from a star of mass $m$ and metallicity $Z$, $\chi_i$ is the initial mass fraction of element $i$ in the star particle, and $y_i(m,Z)$ is the amount of element $i$ created or destroyed in each process which are stored in lookup tables for easy access during the simulation.  The integral is taken over all masses of stars moving off the main sequence between time $t$ and $t+\Delta t$.  

Given that our fiducial SNIa yields are independent of progenitor mass and metallicity, the mass returned by this process is implemented simply for each single stellar population star particle born at time $t_0$ as:
\begin{equation}
M_i(t) = y_i N_{\rm SNIa}(t,t+\Delta t) = y_i \bigintssss_t^{t+\Delta t} g(t' - t_0) dt'
\end{equation}
in which the delay time distribution (DTD) is given by:
\begin{equation}
  g(t)=\begin{cases}
    0, & \text{if $t<\tau_{\rm{8 \, M_\odot}}$}.\\
    N_0 \left( \frac{t}{\tau_{8 \, M_\odot}} \right)^{-s} \frac{s-1}{\tau_{\rm{8 \, M_\odot}}}, & \text{if $t\ge \tau_{\rm{8 \, M_\odot}}$ }.
  \end{cases}
\end{equation}
where the power law index $s=1.12$ is taken as our fiducial value \citep{maoz2012}, the normalization constant is $N_0 = 1.3 \times 10^{-3} \, \rm{M_\odot^{-1}}$ \citep{maoz2012}, and $t\ge \tau_{\rm{8 \, M_\odot}} = 40 \, {\rm Myr}$ is the delay time, here corresponding to the main sequence lifetime of an $\sim 8 \rm{M_\odot}$ star, the upper mass limit for the uncertain progenitors of SNIa.  Following \citet{marinacci2014}, the upper limit for the integration of the DTD has been updated from infinite to the Hubble time (13.79~Gyrs) for consistency with the observations of \citet{maoz2012}. Once again, $y_i$ is the amount of element $i$ created in SNIa events which is stored in look up tables.

Several aspects of the metal enrichment and tracking calculations performed in IllustrisTNG differ from those used in Illustris. An updated second order metal advection scheme is employed~\citep[see Section 2.2 of][for details]{methodspaper}, which improves numerical accuracy during the tracking gas-phase metals within the IllustrisTNG simulation, with only minor increases in computational overhead.  In addition, the yield tables used to calculate $y_i$ for elements ejected by SNIa, SNII and stellar winds have been updated such that the abundances of the 9 explicitly tracked elements more closely follow the ratios found in the Milky Way \citep{nomoto1997,portinari1998,kobayashi2006,karakas2010,doherty2014,fishlock2014}.

A depiction of the most important changes in the revised yield tables is given in Figure~\ref{fig:yields} which shows the mass lost from a simple stellar population over a Hubble time. The top panel shows the ejecta mass from SNIa for all tracked elements heavier than He for a single stellar population, divided by the total mass of the star particle.  Only one line is shown as the SNIa yields are assumed to be independent of progenitor mass and metallicity. The top middle and bottom panels of Fig.~\ref{fig:yields} show the SNII and AGB IMF-weighted ejecta mass fractions for all explicitly tracked elements for a single stellar population, once again, normalized by the total mass of the star particle. 
Because yields for core collapse and stellar wind ejecta depend both on mass and metallicity they are shown weighted by the IMF implemented in IllustrisTNG \citep{chab2003} for the various tabulated metallicities.  It is worth noting that these IMF-weighted yields are estimates of computed ejecta. Ultimately the age of a particular star particle will determine which progenitor masses are dominating the mass loss at any given time.

Overall, the changes from the original Illustris yields predominately affect the SNII yield tables and lead to an increase in neon and magnesium, and to a decrease in carbon, nitrogen, and silicon \citep[as depicted in Figure 1 of][]{methodspaper}. All other elements are hardly changed.

We supplement the monitoring of the individual elements with a tracking of their production channels.  In the IllustrisTNG model, we explicitly follow the total metals produced by SNII, SNIa, and AGB stellar winds, along with the iron produced by the individual SNII and SNIa channels. The initial values of all metal mass fractions and ejecta source tags are set to $10^{-10}$ to suppress numerical advection errors that could otherwise arise when one attempts to track tiny values close to the machine precision. To minimize influences of this lower floor on our results, we limit our discussion to particles and cells with mass fractions above this initial value, $\chi \geq 2.5 \times 10^{-10}$. This results in reliable magnesium to iron ratios down to [Fe/H]$\gtrsim -5.0$ and [Eu/Fe]$\gtrsim 3.25$.\footnote{Throughout the paper we use the common nomenclature $\mbox{[X/Y]} = \log_{10}(N_{\rm X}/N_{\rm Y}) - \log_{10}(N_{\rm X}/N_{\rm Y})_\odot$ where $N_i$ is the number density of species $i$.}  
Raising the value of this mass fraction limit has a negligible effect on the results discussed here. \\
\\
\\
\\
\\

\subsection{Tracking the r-process in IllustrisTNG}

In IllustrisTNG we explicitly follow the r-process production from neutron star -- neutron star (NSNS) mergers, and pay particular attention to the europium ejected from these infrequent events. Similarly to the SNIa implementation, the time dependence of injection is modeled by a DTD, which is approximated as a power law \citep{piran1992,kalogera2001} with index 1.12 for $t > t_{\rm cut}$ and zero otherwise.  The power law index, though generally not well constrained \citep{behroozi2014}, is thought to be close to unity because this is a natural consequence for scenarios in which the total delay time is dominated by the final gravitational inspiral \citep[e.g.,][]{piran1992}.  We take a value of 1.12 such that our DTD for NSNS follows the overall shape of the DTD for SNIa.   In line with observations and other theoretical treatments, we take the cut-off time, $t_{\rm cut}$, to be 100~Myrs \citep{fryer1999,bel2006,shen2015}.

\begin{figure*}
\centering
\includegraphics[width=0.99\textwidth]{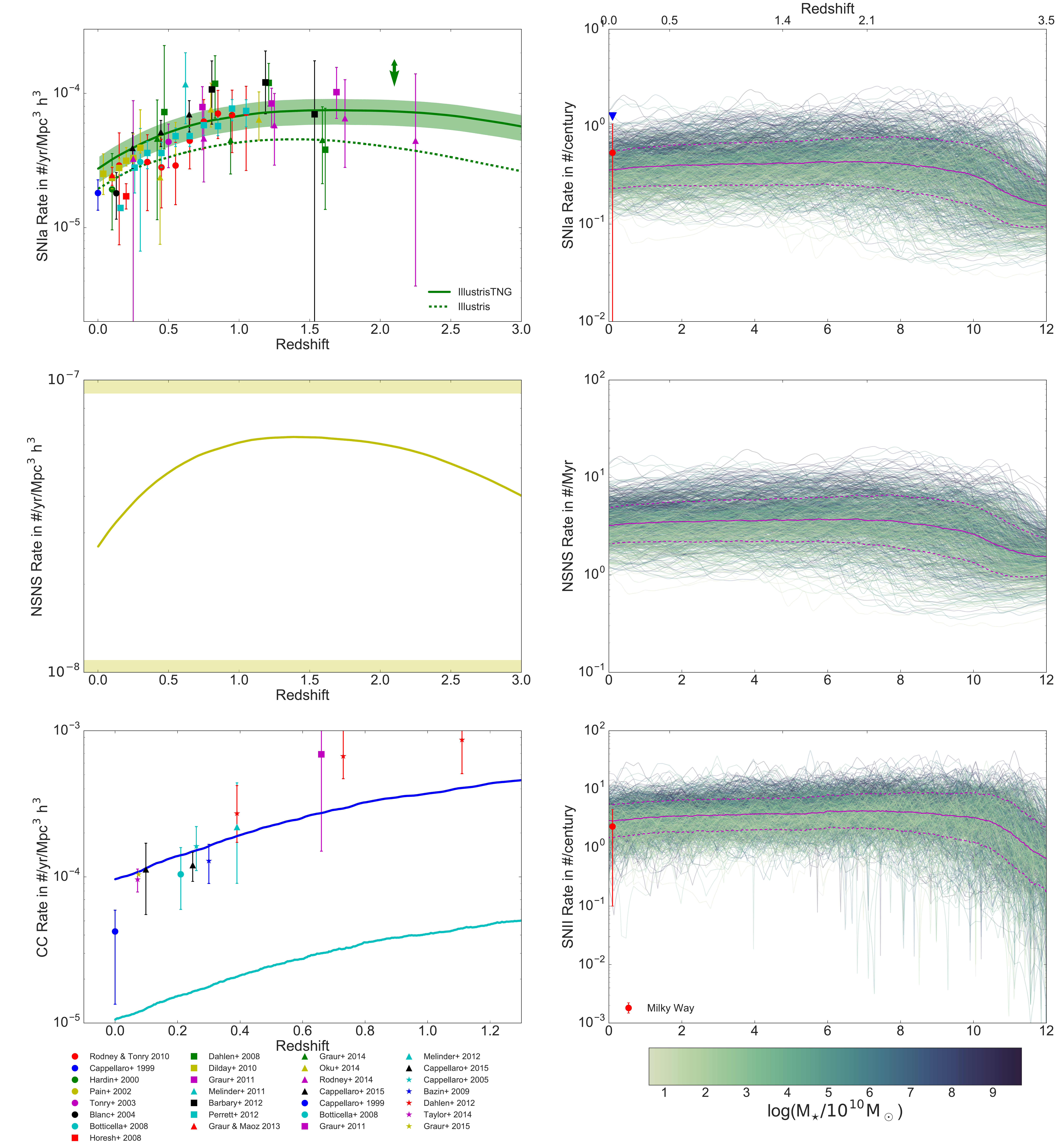}
\caption{
\textit{Left column}: Cosmic SNIa, NSNS and core-collapse SN rates.  
The upper dark blue line in the core-collpase plot is the rate for solar metallicity supernovae, while the lower light blue plot is for zero metallicity. Both SNIa and CC observed rates are from \citealt{cappellaro1999, botticella2008, graur2011, melinder2011, sudare}, CC rates from \citealt{cappellaro2005, bazin2009, dahlen2012, taylor2014, graur2015, sudare}, and SNIa rates from \citealt{rodney2010, hardin2000, pain2002, tonry2003, blanc2004, horesh2008, dahlen2008, dilday2010, barbary2012, perrett2012, graur2013, graur2014, oku2014, rodney2014}. For reference, from \citet{abadie2010} we expect rates of $10^{-8}$,  $10^{-6}$,  $10^{-5}$,  and $5 \times 10^{-5} \, {\rm yr}^{-1}{\rm Mpc}^{-3}$ for their $R_{\rm low}$, $R_{\rm re}$, $R_{\rm high}$, $R_{\rm max}$ models, respectively, listed in their Table 4.  The yellow bar covering the majority of the middle panel indicates the range between the lower limit from \citet{abadie2010}, $R_{\rm low}$, and the approximate upper LIGO/VIRGO limit \citep{bel2016}. The spread in the SNIa rate represents the $2\,\sigma$ errors from fitting the normalization of the IllustrisTNG cosmic rate curve to the data points. Finally, the dashed green line is the SNIa rate from Illustris. \textit{Right column}: SNIa, NSNS, and SNII rates for the 864 Milky Way mass haloes in the TNG100 simulation.  Colors track the stellar mass of the galaxy, with the lighter colors representing haloes with smaller stellar masses and smaller deviations of their enrichment histories from their means. The median is shown with solid magenta lines, and one standard deviation is shown with dashed magenta lines. The observed rate of SNII and SNIa in the Milky Way is relatively uncertain \citep[e.g.,][]{li2011} and is indicated here with red dots.  The SNIa rate multiplied by a factor of 3.5 is shown with a blue triangle, a note of relevance for a subsequent discussion (see Figure \ref{fig:small} for effects of elevated SNIa rate on abundances).  }
\label{fig:cosmicRatesmwRates}
\end{figure*}

Comparisons between our simulated cosmological SNIa rate, NSNS merger rate, SNII rate and observations are shown in the left panel of Figure~\ref{fig:cosmicRatesmwRates}.  As the NSNS merger rate is relatively poorly constrained we caution that some works suggest it may be as high as $5\times 10^{-5} \,  h^3 {\rm yr^{-1} Mpc^{-3}}$ \citep{abadie2010}, several orders of magnitude above our cosmic rate.

\begin{figure*}
\centering
\includegraphics[width=0.8\textwidth]{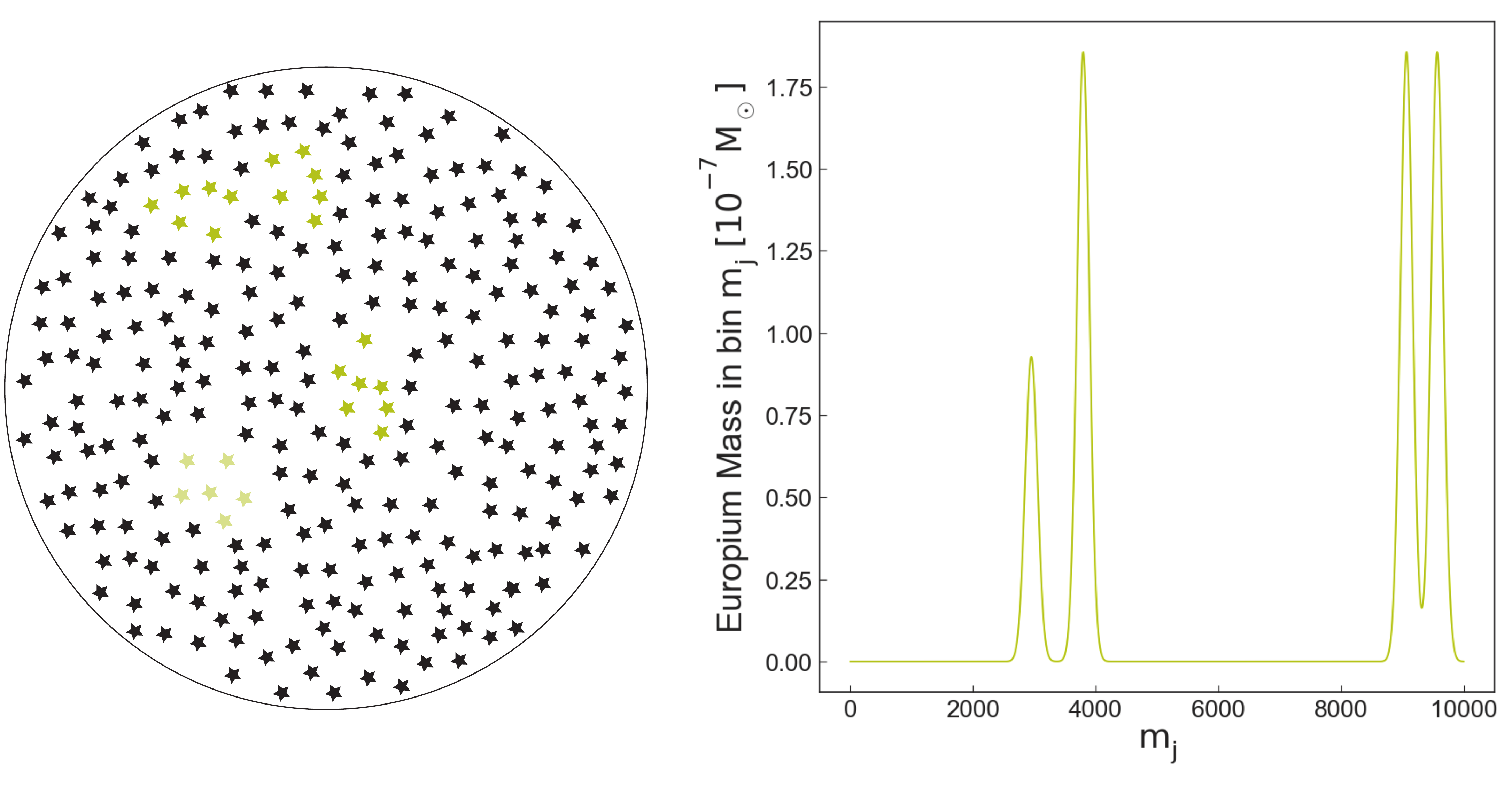}
\caption{An illustration of the sub resolution post-processing method: distribution of NSNS mergers in sub stellar particle mass bins.  The left side sketches an {\small AREPO} star particle broken up into sub-mass bins, and the right side depicts the mass in europium distributed in the sub-mass bins with the total amount of NSNS merger mass in this star particle equaling that of 3.5 NSNS mergers.  The mergers are distributed at sub-mass bins number $\approx 3000, \, 3750, \, 9000, \, 9500$.  Note that the ``fractional'' NSNS merger material is at a lower overall normalization (depicted as fainter stars on the left, submass bin 3000 on the right).}
\label{fig:subgrid}
\end{figure*}

The mass loss rate of r-process material in the Milky Way can be estimated from the r-process mass ejected per NSNS merger, ${M_{\rm rp}}$, the rate of SNIa in the galaxy, $\gamma_{\rm{SNIa,MW}}$, and the rate of NSNS mergers, $\gamma_{\rm{NSNS}}$ as:
 \begin{equation}
 \dot M _{\mathrm{rp}} \approx 10^{-7} \left[ \frac{ {\rm M}_\odot }{ \mathrm{yr} } \right] \left( \frac{ {M_{\rm rp}} }  { 0.05 M_\odot } \right) \left(  \frac{ \gamma_{\mathrm{SNIa,MW}} }{5 \times 10^{-3} \mathrm{yr}^{-1} }  \right)  \left(  \frac{ \gamma_{\mathrm{NSNS}} / \gamma_{\mathrm{SNIa}} }{10^{-3}}  \right) .
 \end{equation} 
 Using values consistent with both observational and theoretical studies of r-process enrichment, ${M_{\rm rp}} = 0.05 \, M_\odot$ \citep{rosswog1999,metzger2010,roberts2011,baus2013,piran2013,grossman2014,just2015,err2015}, $\gamma_{\rm{NSNS}}/\gamma_{\rm{SNIa}} = 10^{-3}$, and $\gamma_{\rm{SNIa,MW}} \approx 5 \times 10^{-3} \rm{yr}^{-1}$ \citep{crocker2016}, we recover the rate of ${\dot{M}_{\rm rp}} \approx 10^{-7} \, {\rm M}_\odot {\rm yr}^{-1}$, which is broadly consistent with observations~\citep{cowan2004, sneden2008}. To obtain a measure of europium, we assume it is produced in solar proportions with respect to the total mass of r-process ejecta, $M_{\rm rp}$, such that $M_{\rm Eu}/M_{\rm rp} = 9.3 \times 10^{-4}$ \citep{sneden2008}.

In IllustrisTNG, the NSNS DTD is implemented stochastically in time to determine how many mergers (if any) have occurred in each star particle as a function of its age. Any NSNS merger material created by an event is then added to the surrounding gas particles in the same manner as all other actively followed species~\citep{vogelsberger2013, methodspaper}.
\\
\\
\\
\\

\subsubsection{Subgrid Distribution of NSNS Merger Material}

Neutron-rich material from NSNS mergers is expected to be incorporated into approximately $10^3-10^4 \, {\rm M}_\odot$ of the ISM into which it is deposited, though this number depends on the ambient density and turbulent velocity, and the energetics of the ejecta \citep{cioffi1988, vandevoort2015, macias2016}. 
Because our gas resolution is larger than this cooling mass, and stars forming from this gas will inherit its uniform value of NSNS ejecta mass, we re-sample this mass to account for unresolved discreteness. Therefore, we assume that when a star particle formed that sub-components of the mass of that stellar population would have sampled a highly fluctuating NSNS-ejecta mass field. To implement this in practice, we break up each star particle into a set of smaller mass particles in post processing
and distribute the mass from each NSNS over these higher resolution bins, a process illustrated qualitatively in Figure~\ref{fig:subgrid}. In post processing, we distribute the material from each NSNS merger in each star particle with individual Gaussians, centered in a sub-mass bin, $m_i$, over all other sub-mass bins, $m$, using a width of $\sigma = 10^4 \, {\rm M}_\odot$. For each star particle $P$, the NSNS ejecta distribution over all sub-mass bins, $m$, takes the form
\begin{equation}
g_P(m) = \frac{A}{\sqrt{2 \pi (\sigma/M_b)^2}}\, {\rm e}^{-{\frac{(m-m_i)^2}{2\,(\sigma/M_b)^2}}} .
\label{eq:gaussian}
\end{equation}
Here the dispersion $\sigma$ is normalized by the mass in each sub-mass bin, taken to be $M_b = 100 \, {\rm M}_\odot$.  The randomly assigned sub-mass bin $m_i$ is the bin where the $i$-th NSNS merger ejecta is assumed to be incorporated. The normalization constant $A$ of the function $g_P(m)$ is chosen such that the total mass summed over all of the sub-mass bins adds up to the total NSNS mass in each star particle, i.e. the total mass in the cell $\Sigma_{\rm NSNS \, \, mergers} \int g_P(m)\, {\rm d} m$  equals the total NSNS mass in each star particle $P$. Here we assume that the parameters of this model - $M_b$, $\sigma$ - do not change as a function of redshift. 

The subgrid prescription described here ignores the effects of natal kicks on displacing NSNS mergers from their birth position.  When natal kicks are significant such displacement is non-negligable \citep{bloom1999,fryer1999,bel2006,zemp2009,kelley2010,behroozi2014} and if the NSNS merger travels to a region where gas densities are lower, the cooling mass is larger, thereby spreading any r-process ejecta over a wider region, potentially smoothing out any large europium enhancements \citep{macias2016,saf2017} .

In addition, the current implementation of the subgrid prescription assumes a conservative amount of mixing between the neutron-rich NSNS material and the pristine gas in the time between enrichment events and the formation of stars.  This assumption relies on the eddy turnover time scale to be large compared to the delay time between enrichment and star formation.  When the eddy turnover time for a cell is approximated from the symmetric part of the shear tensor of surrounding cells \citep[e.g.][]{sar2017}, we find at high redshift the majority of the gas cells should form stars before they are well mixed.
While the assumption of minimal mixing is a good approximation at high redshift and for most of the cells at low redshift, a fraction of the cells are well mixed at low redshifts.  We caution that given our limited number of saved snapshot files we can only provide estimates for these timescales.  \\
\\
 The motivation, detailed properties of the implementation of this subsampling, and the effects on our analysis when the cells are well mixed are discussed more fully in Section~\ref{section:discussion}.

\begin{figure}
\centering
\includegraphics[width=0.48\textwidth]{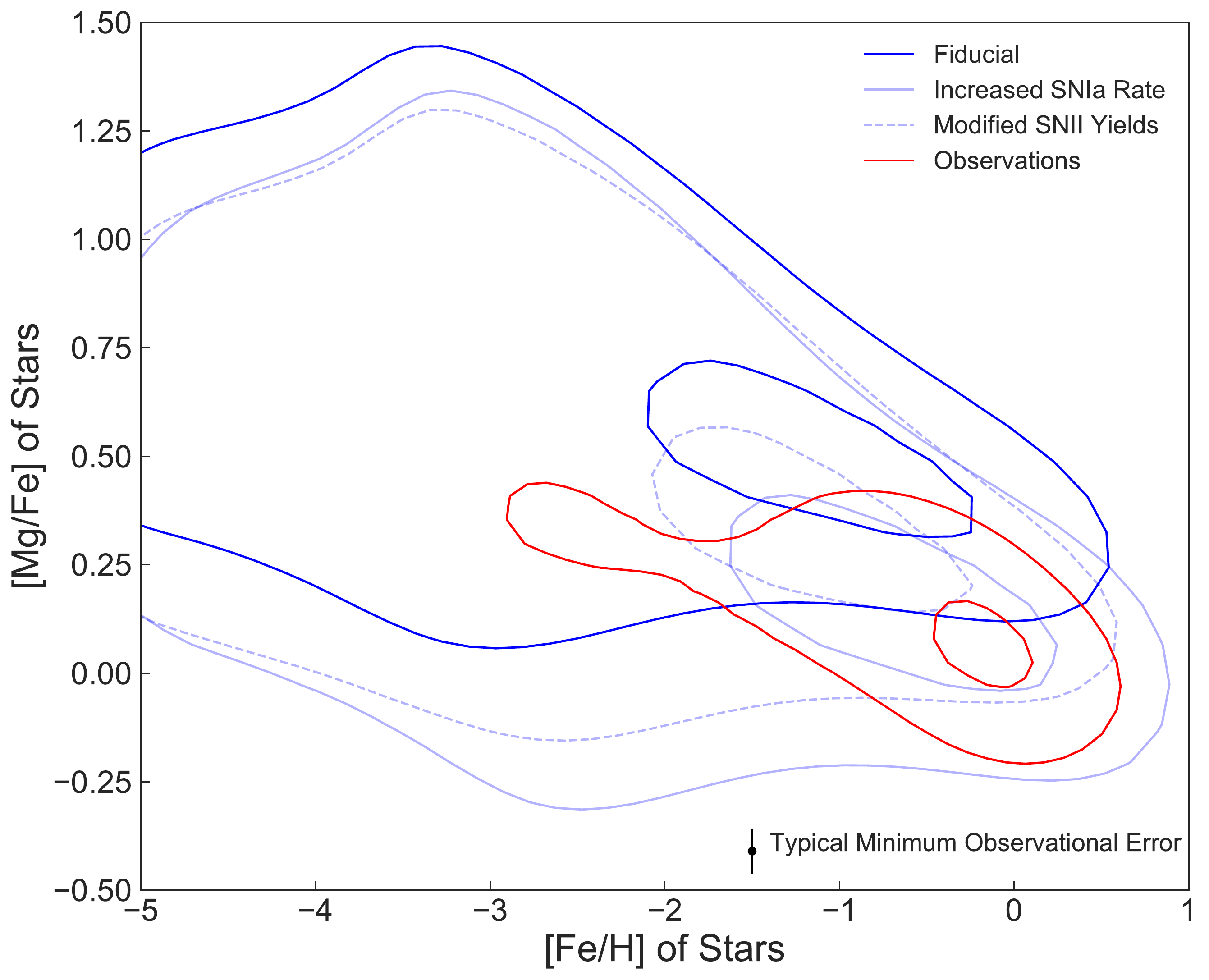}
\caption{The [Mg/Fe] ratio in stars of both simulated IllustrisTNG Milky Way sized galaxies (blue contours) and that observed in the Milky Way (red contour) \citep{APOGGEE_R13, gratton2003, reddy2003, cayrel2004, venn2004, bensby2005, reddy2006, lai2008, bonifacio2009, and2010, frebel2010, ad2012, ish2012, cohen2013, yong2013, bensby2014, hinkel2014, roed2014, jacob2015}.   The simulated data contains all stars bound to each Milky Way halo.  Abundances have been renormalized to the solar values of \citet{asplund2005}. The [Mg/Fe] ratio of simulated star forming Milky Way-sized galaxies (dark blue lines) is higher than observed (red contours).  Besides complications in the comparison of observational and simulated abundance ratios (see Figure~\ref{fig:large}), other depicted explanations for the discrepancy are a low rate of SNIa or incorrect SNII yields for Mg and/or Fe. The [Mg/Fe] discrepancy is alleviated in the Milky Way-sized galaxies if the SNIa rate is increased by a factor of 3.5 in post processing (light blue solid lines; see Figure \ref{fig:cosmicRatesmwRates}) or if the SNII Mg and Fe yields are modified (light blue dashed lines, see text for further details). Contours of each color are the 50\% and 5\% kernel density estimation levels.}
\label{fig:small}
\end{figure}

\subsection{The IllustrisTNG Suite} \label{section:illustris}
 
To analyze the distribution of r-process material in Milky Way mass haloes, we utilize the IllustrisTNG suite of simulations, the successor of the Illustris project~\citep{vogelsberger2015a, vogelsberger2015b, genel2015b, sij2015},  which employs the moving mesh code {\small AREPO} \citep{arepo} to evolve the gravitational and hydrodynamical equations combined with the galaxy formation physics module described extensively in \citet{weinberger2017} and \citet{methodspaper}. For the sake of conciseness, we here only briefly summarize the most relevant methodology used in the IllustrisTNG simulations, making specific note of our star formation, cooling, metal creation and return implementations. For full details on the numerical methods and subgrid physics prescriptions we refer the reader to the work of \cite{weinberger2017} and \citet{methodspaper}.

The IllustrisTNG simulations adopt the Planck intermediate results for the cosmological parameters, $\Omega_{\rm M} = 0.3089$, $\Omega_{\rm \Lambda} = 0.6911$, $\Omega_{\rm b} = 0.0486$, $h = 0.6774$, and $\sigma_{\rm 8} = 0.8159$ \citep{planck2016}. Our simulations were all started at redshift $z=127$, and produced 100 time-slice outputs during the time evolution down to the present epoch. The results presented here are based on the TNG100 simulation, a $75\,h^{-1}{\rm Mpc} \simeq 100\,{\rm Mpc}$ periodic box with the initial conditions containing 1820$^3$ dark matter particles and the same number of gas cells. The average gas-cell and star particle mass is then approximately $1.4 \times 10^6 \, {\rm M}_\odot$, while the dark matter particle mass is $7.5 \times 10^6 \, {\rm M}_\odot$.  The dark matter and stellar softening length is approximately 740~pc at $z=0$, while the gas softening lengths are adaptive with a minimum of 185 comoving parsecs. Besides TNG100, the simulation suite of IllustrisTNG also encompasses a larger box of $\approx 300\,{\rm Mpc}$, referred to as TNG300, and a smaller box of size $\simeq 50\,{\rm Mpc}$ (TNG50, still ongoing at the time of this writing). Each of the boxes has also been run at lower resolutions and as dark matter-only versions, which are however not used in the present study.   A summary of the TNG100 run parameters are presented in Table \ref{table:specs}.  While the majority of this paper relies on the results from the highest resolution box, TNG100-1, we include the specifications for a lower resolution run, TNG100-2, as we make use of this simulation to test the convergence of our models. 

The TNG100 simulation accounts for radiative gas cooling~\citep{katz1996} including contributions from metal lines~\citep{vogelsberger2013}, which is dependent on the gas density, temperature, metallicity, and redshift. The effects of self-shielding, the UV background \citep{faucher2009} and radiative modifications to the cooling rates due to AGN are taken into account \citep{vogelsberger2013, methodspaper}.  Gas is transformed into stars via a subgrid star formation prescription, which assumes a star formation density threshold of $n_{\rm H} \simeq 0.1 \, {\rm cm}^{-3}$ and a gas consumption timescale of $\sim$2.27~Gyrs at this density~\citep{kennicutt1998,springel2003}.  Star particles are formed by stochastically sampling the estimated star formation rate and, once formed, each star particle represents a simple stellar population which follows a~\citet{chab2003} IMF. Such a star particle retains the metallicity and metal abundances of the gas cell from which it was formed. 

The gas content of a galaxy is strongly impacted by the implemented feedback processes, most notably galactic winds, which transport substantial quantities of metal rich material into the regions surrounding galaxies \citep{methodspaper}. The mixing processes induced by the resulting complicated flows are treated naturally by our mesh code, albeit with lower rates of numerical diffusion than fixed grid codes \citep{springel2010}. The use of {\small AREPO} therefore allows us to forgo schemes required in other studies for sub-resolution metal diffusion \citep{shen2015}, or for ``sticking'' abundances to gas particles once enriched and which is associated with an underestimate of the diffusion of elements in SPH codes which offer no numerical diffusion \citep{hopkins2017,hubber2013,vandevoort2015}.

\begin{table}
\caption{Main numerical parameters of the two realizations of the TNG100 Simulation analyzed in this paper.   }
\label{table:specs}
\begin{tabular}{lcc}
\hline \hline
Parameter & ${\rm{TNG100(-1)}}$ & ${\rm{TNG100-2}}$ \\
\hline
$N_{\rm{gas}}$  & ${\rm{1820^3}}$ & ${\rm{910^3}}$\\
$N_{\rm{DM}}$ & ${\rm{1820^3}}$ & ${\rm{910^3}}$\\
$N_{\rm{TRACER}}$ & ${\rm{2 \times 1820^3}}$ & ${\rm{2 \times 910^3}}$ \\
$m_{\rm{baryon}}$ [${\rm{10^6 \, M_\odot}}$] & ${\rm{1.4}}$ & ${\rm{11.2}}$ \\
$m_{\rm{DM}}$ [${\rm{10^6 \, M_\odot}}$]& ${\rm{7.5}}$ & ${\rm{59.7}}$ \\
$\epsilon_{\rm{gas,min}}$ [${\rm{kpc/h}}$]& ${\rm{0.25}}$ & ${\rm{0.5}}$ \\
$\epsilon_{\rm{z=0,(DM,stars)}}$ [${\rm{kpc}}$]& ${\rm{1.48}}$ & ${\rm{2.95}}$ \\
\hline
\end{tabular}
\end{table}

\subsection{Milky Way galaxies in TNG100} \label{section:mws}

In order to compare our simulated abundance ratios with those observed in the Milky Way (MW), we begin by defining selection criteria that determine which galaxies we use as Milky Way analogs. First, we select for galaxies with a total halo mass, $M_{\rm 200, crit}$,  in the range $(0.6 - 2) \times 10^{12} \, {\rm M}_\odot$ \citep{eadie2015, mcmillan2017, zar2017}, which results in approximately 2000 galaxies. Next, we require our MWs to reside close to the SFR-${M_\star}$ main sequence~\citep[e.g.][]{noeske2007}. This naturally filters out red galaxies with very low star formation rates, leaving us with $\sim$1400 galaxies. Finally, we require our MW-like galaxies to be ``disky''.  This is quantified by having a large fraction of stellar mass with a high value of the circularity parameter, where the circularity parameter for an individual star, $\epsilon = J_z/J(E)$, measures the angular momentum of the star about the axis of symmetry in units of the maximum angular momentum possible at the same orbital energy $E$. For a circular orbit in the plane perpendicular to the symmetry axis, one has $\epsilon =1$ \citep{marinacci2014}.

A commonly employed definition of the ``disk'' in a simulated galaxy is the collection of stars with $\epsilon > 0.7$. Thus, the fraction of stars with high circularities, $f_{\epsilon > 0.7}$, determines how much of a galaxy's stellar population resides in the disk component \citep{marinacci2014, genel2015}. We split our population of galaxies along the median of our $f_{\epsilon > 0.7}$ distribution, and require our Milky Way selection to come from galaxies with fractional circularities larger than this median.

The final selection of 864 galaxies forms the basis of the discussion of Milky Way-sized galaxies in this paper. Our criteria still allow for a variety of enrichment histories, as depicted by the range of SNIa, NSNS, and SNII enrichment histories for the individual Milky Way-size galaxies in the right column of Figure~\ref{fig:cosmicRatesmwRates}, with haloes hosting less massive galaxies having somewhat less diverse enrichment histories.

 \begin{figure*}
\centering
\includegraphics[width=1.0\textwidth]{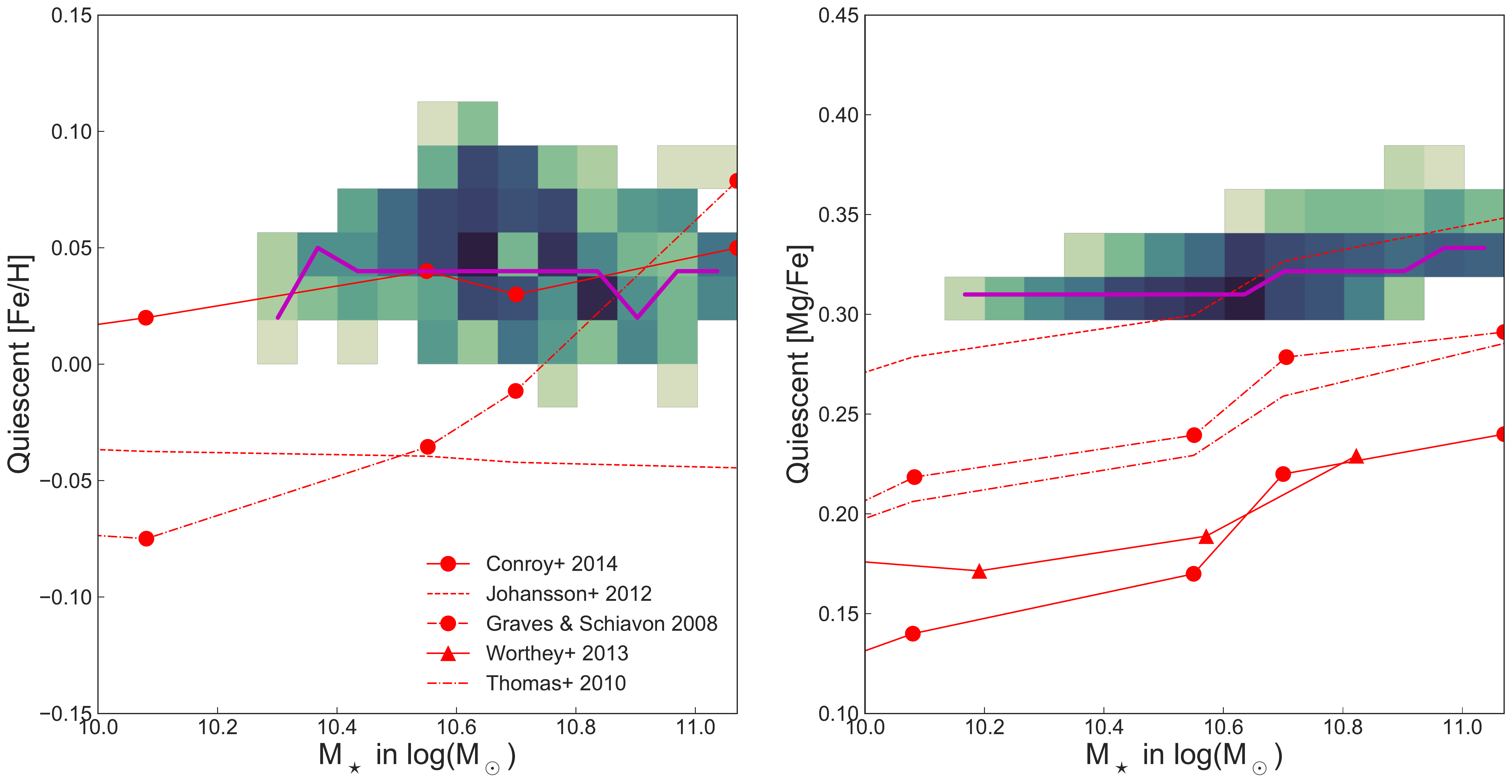}
\caption{The scales of abundance ratios of [Fe/H] and [Mg/Fe] 
over a wide range of quiescent galaxy masses.  Here, simulated points are the galaxy-averaged stellar components within the stellar mass half radius stacked in each stellar mass bin and observed points are the result of stacked spectra of several galaxies in each stellar mass bin.  Only centrals and galaxies which have not recently undergone a burst of star formation ($\rm{M}_\star(\rm{age} \le \rm{1\, Gyr})/M_\star < 20\%$ are included).
Observations of quiescent galaxies are denoted by red points, lines and triangles \citep{graves2008, thomas2010, joh2012, worthey2013, conroy2014}. All abundances have been renormalized to the solar values of \citet{asplund2005}. Magenta lines and green histograms show the simulated ratios of the r-band luminosity weighted medians of stellar abundances in galaxies with low specific star formation rates, $\rm{sSFR \le 10^{-11} \, yr^{-1}}$.  Histograms are derived with stellar mass bins of $0.071\log{\rm{M}_\odot}$, and bins with less than 10 galaxies are discarded.  The range from light to dark colors contains 11-38 and 11-115 galaxies for a total of 1107 and 1267 galaxies for the [Fe/H] and [Mg/Fe] histograms, respectively. The [Mg/Fe] ratios for quiescent galaxies are  higher than observed as well, which is depicted in the right panel.  Besides complications in the comparison of observational and simulated abundance ratios \citep[e.g.][]{guidi2016}, another possible explanation for this discrepancy is a too low rate of SNIa (see Figure \ref{fig:small}) or a too high production of Mg from SNII (see text). }
\label{fig:large}
\end{figure*}

\begin{figure*}
\centering
\includegraphics[width=1.0\textwidth]{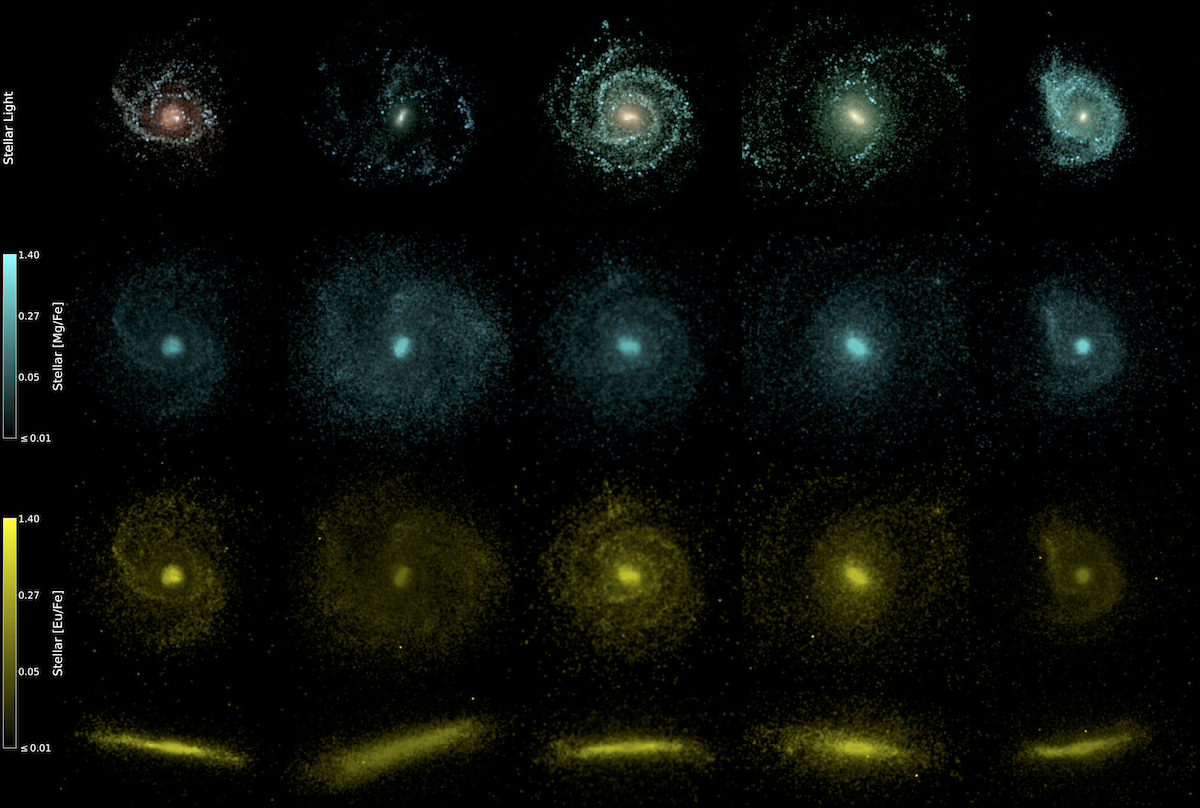} 
\caption{Morphology of several Milky Way mass galaxies along with the distribution of their [Mg/Fe] and [Eu/Fe] abundance ratios at $z=0$. The panels on top show mock images of the simulated galaxies from the stellar light distributions (g-, r-, and i-bands).  Middle row images show a mass weighted map of the maximum [Mg/Fe] in any star particle.  Bottom row images show the maximum [Eu/Fe] in each star particle for an individual galaxy in both face on and edge on images.  Only a few particles with high [Mg/Fe] enhancements appear in any galaxy while several galaxies have highly enhanced [Eu/Fe].  In addition, the overall spread of [Eu/Fe] is larger than that of [Mg/Fe] for any given galaxy as shown by the relative saturation of the yellow images.}
\label{fig:mwimages}
\end{figure*}

\begin{figure*}
\centering
\includegraphics[width=0.8\textwidth]{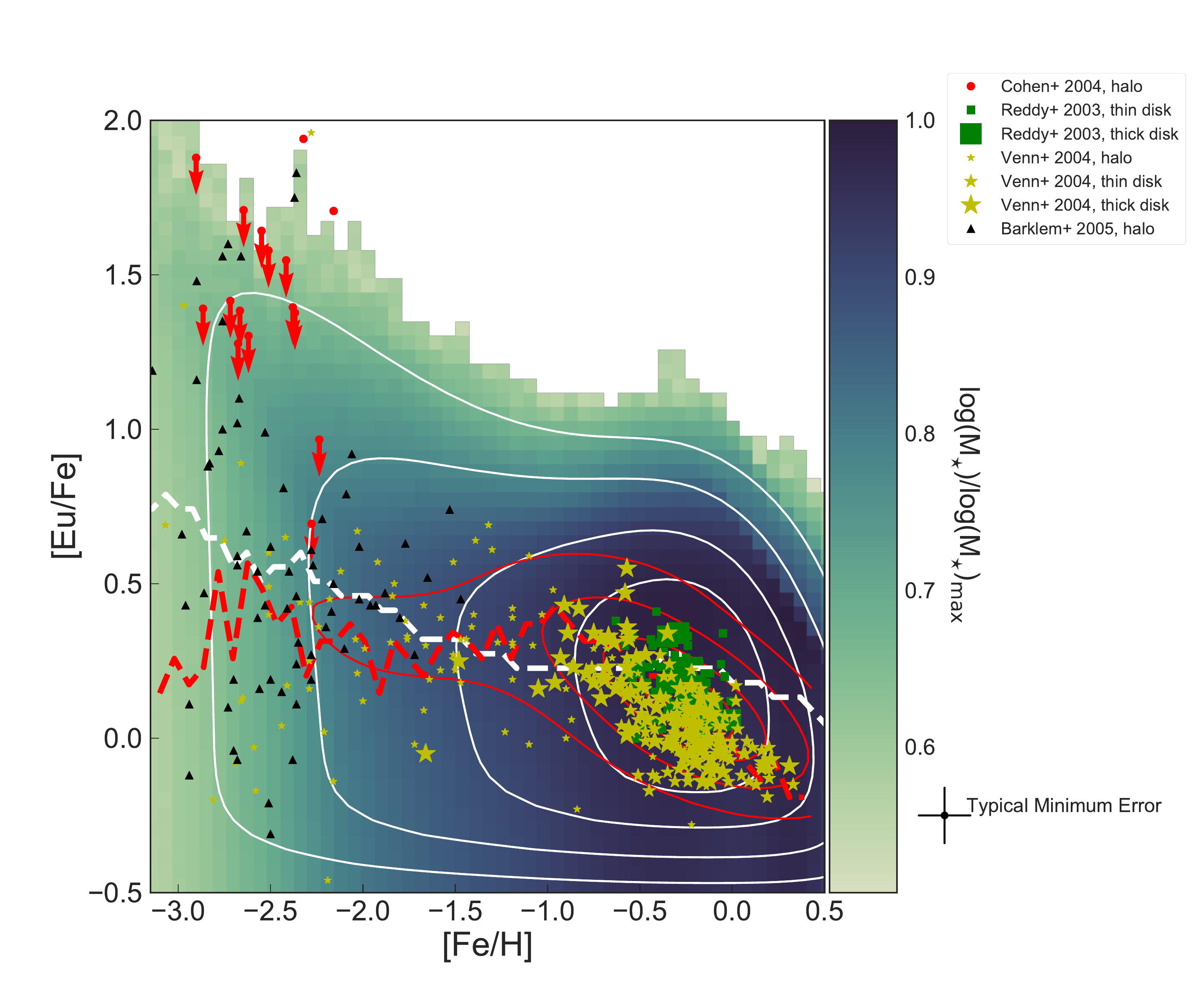}
\caption{Distribution of [Eu/Fe] vs. [Fe/H] for all sub-mass bins of all bound star particles of our simulated Milky Way-like galaxies, binned in a $50 \times 50$ histogram.  Red (white) solid lines denote the 75\%, 40\%, 25\% and 10\% contour levels of the observed (simulated) density of data points while dashed lines show the medians of the distributions in each [Fe/H] bin. Observations denoted with red contour levels are compiled by the NuPyCEE package \citep{nupycee} and include halo, thick, and thin disk stars \citep{reddy2003, venn2004, bensby2005, reddy2006, aoki2008, lai2008, roed2009, frebel2010, hansen2012, ish2012, hinkel2014, roed2014, jacob2015, batt2016}. Observationally, Milky Way halo stars dominate the contribution to the [Eu/Fe] ratio at low [Fe/H], as shown by the highlighted observations of \citet{cohen2004, venn2004}.  In contrast, thin and thick disk stars make up the majority of the contribution at higher metallicities \citep{reddy2003, venn2004}. In the \citet{reddy2003} observations, membership to the thick disk is denoted by having a line of sight velocity $> -40.0 \, {\rm km\,s^{-1}}$, while membership in any component of the \citet{venn2004} observations is based on their component membership probability being larger than 0.5.  Other sources have selected for halo stars explicitly \citep{cohen2004, barklem2005}.  Here, and in following [Eu/Fe] vs. [Fe/H] plots, lower confidence interval KDE lines are only approximations to the total [Eu/Fe] vs. [Fe/H] distribution as several cuts have been applied to the included star particles due to numerical effects (see text for details).}
\label{fig:eufeWithobs}
\end{figure*} 

\begin{figure*}
\centering
\includegraphics[width=1.0\textwidth]{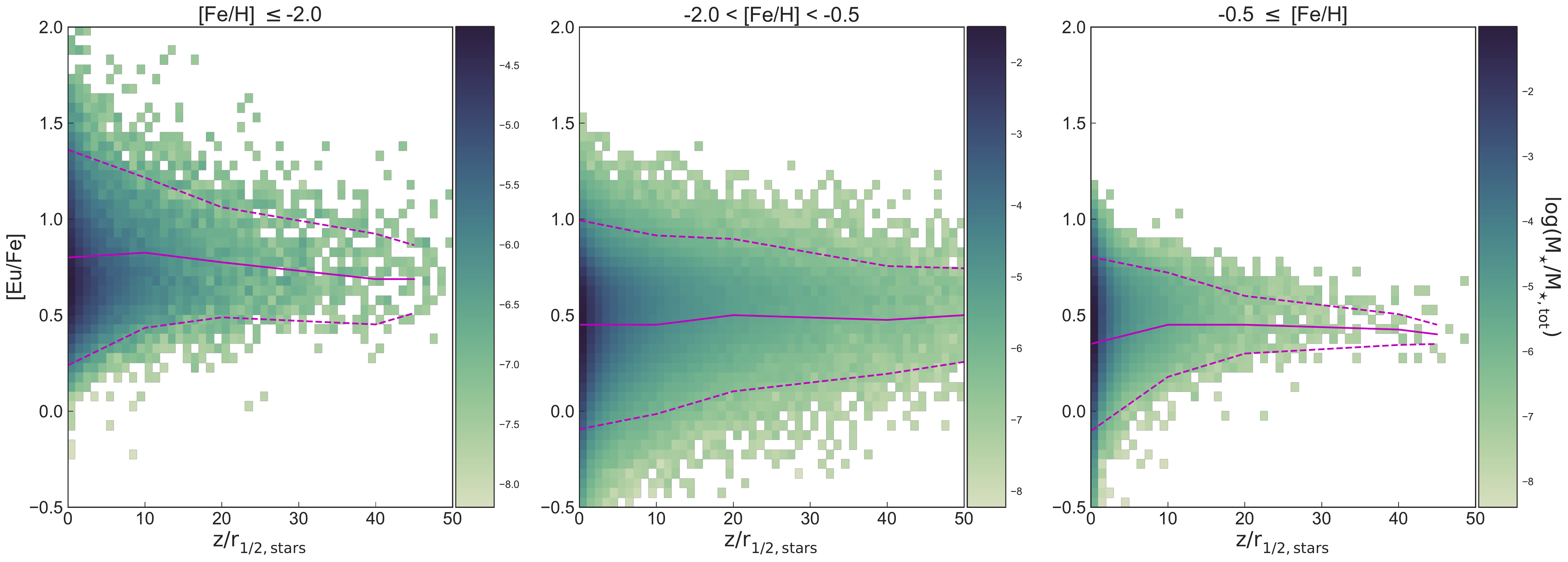}
\caption{Spatial distribution of the maximum [Eu/Fe] in each star particle for several [Fe/H] bins, showing the distribution of metal poor and metal rich stars in both their [Eu/Fe] enrichment and location with respect to the disk of the galaxy.   Median and one standard deviation are shown with solid and dashed magenta lines, respectively. The z-axis of a disk is defined as the plane perpendicular to the stellar angular momentum vector.  Lower and intermediate metallicities tend to have larger [Eu/Fe] enhancements and show a preference to exist in both the plane of the galaxy and further out in its halo than their higher metallicity counterparts.  Here, the histogram colors give the total mass in a bin over all the summed stellar mass from all Milky Ways in TNG100.}
\label{fig:spatialDist}
\end{figure*}

 \begin{figure*}
\centering
\includegraphics[width=1.0\textwidth]{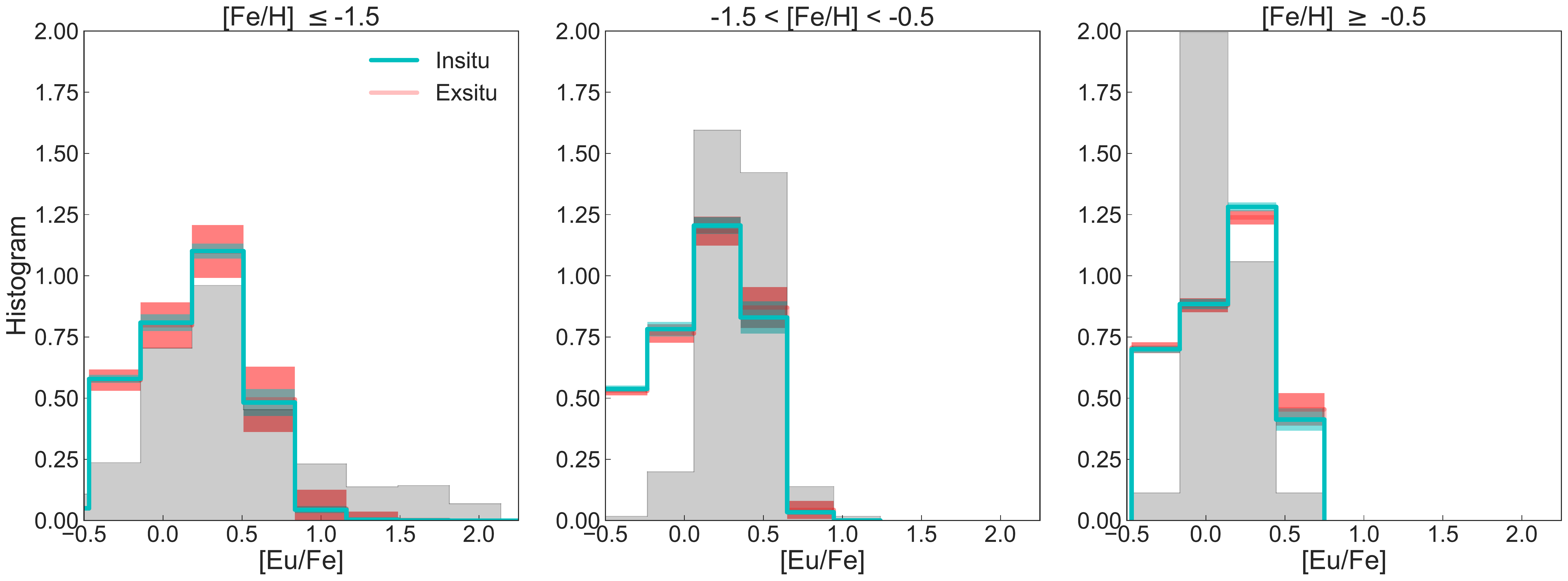}
\caption{Distributions of stellar particles' sub-mass distributions of [Eu/Fe] in different [Fe/H] bins averaged over the full sample of haloes (thick lines), color coded by the origin of the stellar particles: in-situ (cyan) and ex-situ (red).  Grey histograms are observational data collected with the NuPyCEE package \citep{nupycee}. Bars show the spread for the in-situ and ex-situ histograms, normalized by the total mass in each stellar population.   All histograms have been normalized by their total area.  The simulated distributions generally agree well with observations, though are generally wider in spread and show less evolution with [Fe/H] than the observations (see text for further details).  In addition, there is little difference between the simulated in-situ and ex-situ distributions.}
\label{fig:eufeinex}
\end{figure*}

\begin{figure*}
\centering
\includegraphics[width=1.0\textwidth]{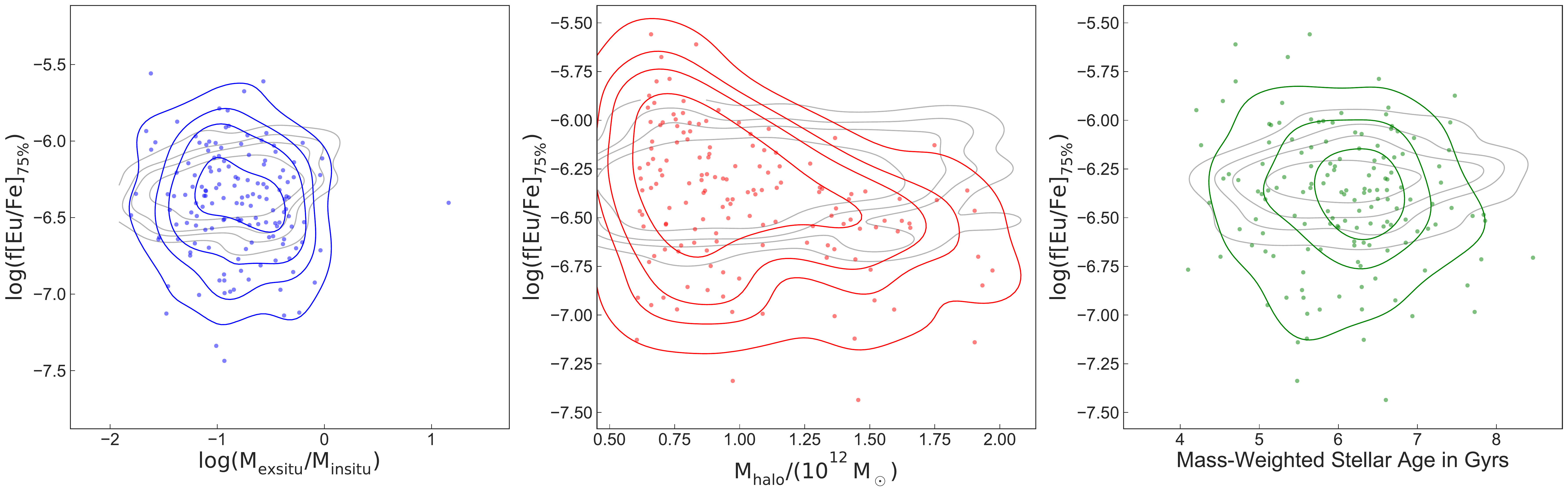}
\caption{Mass fraction of sub-mass star particles in Milky Way-sized galaxies with $\mbox{[Eu/Fe]} \gtrsim 1.3$ with respect to a variety of galaxy parameters.  Each point represents results taken from all star particles bound to a single galaxy. The left panel gives the ratio of the ex-situ to in-situ stellar populations, the middle panel is the total halo mass of the galaxy, and the right-most panel shows the mass-weighted stellar population age.  The contours are based on kernel density estimates from the seaborn Python package, and are meant to guide the eye.  No strong trends are present, with slight trends possible only with halo mass and stellar age. For comparison, lines for the mass fraction of enhanced [Mg/Fe] stars ($\mbox{[Mg/Fe]}\gtrsim 0.97$) are shown as gray contours.  These have been renormalized such that they overlay on the [Eu/Fe] contours (by a multiplication factor of about 8).}
\label{fig:fractions}
\end{figure*}

\begin{figure*}
\centering
\includegraphics[width=1.0\textwidth]{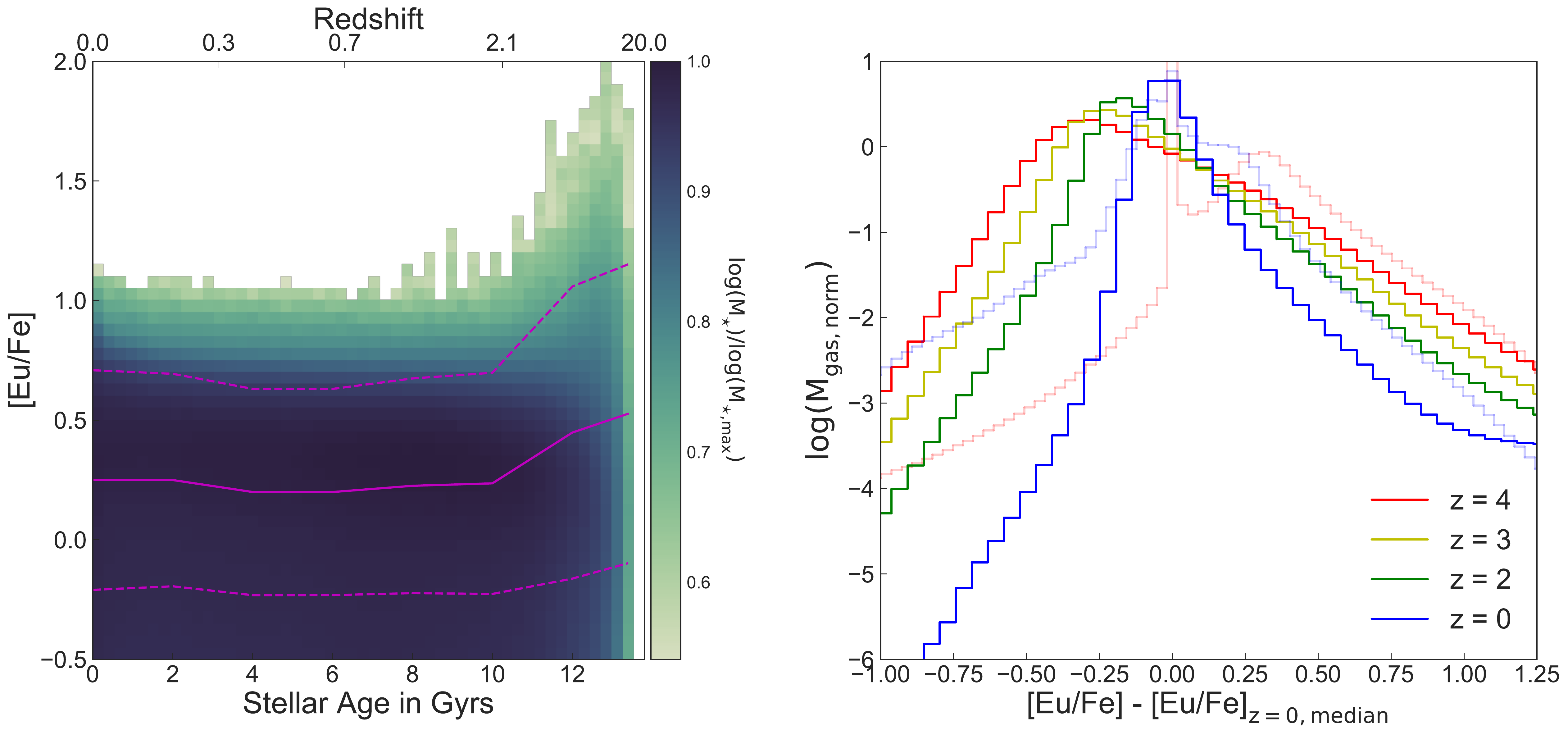}
\caption{Relation between stellar sub-mass [Eu/Fe] enhancement and age (left panel), and the cosmological gas phase distributions of [Eu/Fe] for different redshifts (right panel).  In the left panel, magenta solid and dashed lines show the median and standard deviation of [Eu/Fe] for each stellar age bin.  The stars with the largest [Eu/Fe] tend to reside above the 75\% cut and are born at early redshifts.  Thick lines in the right panel show gas phase [Eu/Fe] both increases and becomes more homogeneous with time.  The [Mg/Fe] in the gas phase shows the opposite trend (light red (blue) lines at $z=4$ ($z=0$)) -- growing less homogeneous with time as they shift from high [Mg/Fe] to lower [Mg/Fe] as SNIa begin to pollute the gas.  All gas phase histograms have been normalized by their total area.  Taken together, these plots indicate that older stellar populations are born from gas with a larger spread of [Eu/Fe] abundances and therefore have a greater chance of forming from highly r-process enhanced gas during $z=2-4$. }
\label{fig:agespread}
\end{figure*}

\section{Observational Abundance Comparisons} \label{section:observations}
 
In this section we compare the [Mg/Fe] and [Fe/H] stellar abundance ratios found in IllustrisTNG against observations. To provide an even-handed comparison between the simulation derived abundance ratios and the observationally derived values, we split our comparison into two parts: (i) a comparison of abundance ratios in Milky Way-sized galaxies in TNG100 with trends seen in our actively star-forming galaxies, and (ii) a comparison of abundance ratios in quiescent galaxies that show rich enrichment histories in these inactive populations as well \citep{graves2008, thomas2010, joh2012, worthey2013, conroy2014}. Specifically, we compare the globally-averaged abundance ratios in the simulated quiescent galaxy sample against stacked SDSS spectra and compare the distribution of [Mg/Fe] and [Fe/H] values for individual stellar populations against  stellar populations observed in the Milky Way. For this initial comparison we consider only magnesium and iron, as europium is only observable in the local stellar population, not in the distant quiescent galaxy populations.

\subsection{Alpha abundances in Milky Way-like simulated galaxies}

Figure~\ref{fig:small} shows a comparison of the [Mg/Fe] ratio observed in stars in the Milky Way (red contours) against the [Mg/Fe] ratio found within simulated Milky Way analogs in IllustrisTNG (dark blue contours), as a function of the iron-to-hydrogen ([Fe/H]) ratio. The observed and simulated metallicity distributions broadly occupy the same space. However, in more detail, the [Mg/Fe] ratio for the simulated MW systems (dark blue contours) is somewhat higher, by $\sim$0.2 dex, than that of observations of individual stars in our own Milky Way (red contours).  While Figure~\ref{fig:small} shows the [Mg/Fe] distribution of stars stacked across all of our simulated Milky Ways, the offset is present in all individual galaxies as well.
The offset is unlikely due to the observational errors. Although the observational measurements of Figure \ref{fig:small} rely on simplistic stellar models, recent work employing more physically realistic non-LTE models with three dimensional stellar atmospheres generally confirms the observed [Mg/Fe] distributions seen in the earlier observational studies~\citep{berg2016}.

One possible origin of this offset could be an underestimate of the average Milky Way SNIa rate. Motivated by the uncertainties in both the observed SNIa rate \citep[e.g.][]{shafter2017} and the normalization of the SNIa DTD \citep{graur2011,graur2013b}, we  recalculate the [Mg/Fe] vs. [Fe/H] distribution assuming an increase of the SNIa rate. This is achieved in practice by identifying the Mg and Fe that was produced by SNIa via the ejecta source tags, and rescaling the SNIa contributions appropriately. This is possible because the SNIa yields are independent of the mass, metallicity, or age of the progenitors, and so we can simply rescale Mg, Fe, and H by the assumed increase/decrease in the SNIa rate.

The modified [Mg/Fe] vs. [Fe/H] distribution assuming a SNIa rate three and a half times higher than that found in the TNG analogs is indicated by the solid light blue lines in Figure~\ref{fig:small}. As SNIa eject most of their mass as iron, the increase in SNIa rates in post-processing predominately affects the [Mg/Fe] distribution by lowering the [Mg/Fe] ratio at higher metallicities. 
While such an increase in SNIa rate is not ruled out by observations, it would result in a Milky Way SNIa rate somewhat higher than observed, as shown with a blue triangle in the top-right panel of Figure \ref{fig:cosmicRatesmwRates}.  

As much of the magnesium production is in SNII, it is natural to also explore the effects of altering the magnesium and iron production in SNII as a means to overcome the discrepancies between the simulated and observed abundance ratio.  Because SNII yields are dependent on both the mass and metallicity of the progenitor, a post-processing alteration to the yields as was done for with the SNIa source tag is not straightforward.  Instead, we perform a coarse estimate of the effects of altered SNII yields on the simulated [Mg/Fe] ratio.  In the dashed blue line of Figure~\ref{fig:small}, we use to SNII source tag to artificially decrease the total mass in magnesium from SNII by a factor of 2, and add this mass into iron.  This change, while large, is within the differences present between different SNII yield generation groups \citep{portinari1998,kobayashi2006}.

 While the application of the SNIa and SNII source tags here are used only to illustrate the effects of changing the uncertain SNIa rate  
 in Milky Ways on their [Mg/Fe] distribution, a fuller discussion of yield variations with different rate and progenitor model assumptions will be the topic of a future paper ({\textcolor{blue}{Naiman et.\ al, in prep}}).

\subsection{Alpha and Iron abundances in quiescent galaxies}

The left panel of Figure~\ref{fig:large} shows a comparison of the galaxy wide [Fe/H] ratio as a function of stellar mass for quiescent galaxies. 
Here our sample of quiescent galaxies includes all those with low specific star formation rate, $\rm{sSFR \le 10^{-11} \, yr^{-1}}$, which is similar to the cuts in line emission in observationally defined quiescent galaxies \citep{graves2008}. In addition, to facilitate comparison with observations we further limit our sample in two ways.  To mimic the observational cuts excluding superimposed images \citep{graves2009,peek2010} we select only central galaxies for our sample.  In addition, we avoid selecting recent star burst galaxies, by further limiting our sample to galaxies in which no more than 20\% of the stellar mass is from stars with ages $\le$ 1.0~Gyrs.
The simulated [Fe/H] values are calculated as the r-band luminosity-weighted [Fe/H] average for all stars in the galaxy using the \citet{bruz2003} models. The background two-dimensional histogram in the left panel of Figure~\ref{fig:large} indicates the distribution of [Fe/H] values for the simulated quiescent galaxy population and the magenta solid line indicates the median relation of this histogram.  Bins containing less than 10 galaxies have been excluded from the plotted histograms. For comparison, the red lines indicate observations of [Fe/H] values for quiescent galaxy populations, as indicated in the legend~\citep{graves2008, thomas2010, joh2012, worthey2013, conroy2014}. 
While the simulated [Fe/H] values are relatively flat, we caution here that this result is somewhat sensitive to both bin sizes and exclusion criteria - when less populated bins are included or the bin sizes are smaller there is a slight turn over in [Fe/H] at $M_\star \lesssim 10^{10.4}\, {\rm M}_\odot$.

In comparison to the observations (red symbols and curves), the simulated r-band luminosity-weighted [Fe/H] values are somewhat higher (i.e. $\sim0.05-0.15$ dex) than found for the observed systems, even though we note that different observational analyses differ by up to $\sim$0.1~dex in normalization and return rather discrepant slopes.  In addition to effects of binning and low numbers of simulated galaxies at low stellar masses, these differences can, in part, be influenced by the method used to perform the comparison between the observationally derived and simulated [Fe/H] values \citep[e.g.,][Figure 2 of \citealt{nelson2017}]{guidi2016}.

The right panel of Figure~\ref{fig:large} shows a comparison of the galaxy wide [Mg/H] ratio as a function of stellar mass for quiescent galaxies. As with [Fe/H], simulated [Mg/Fe] values are calculated as the r-band luminosity-weighted [Mg/Fe] average for all stars in the galaxy using the \citet{bruz2003} models. The [Mg/Fe] ratios shown in the right panel for our sample of quiescent galaxies further demonstrate the complex relationship between simulated and observed abundance trends.  The simulated [Mg/Fe] values are gradually increasing with increasing stellar mass and show limited scatter across the full mass range explored in Figure~\ref{fig:large}. 
In contrast, while different observational estimates can differ in their normalizations by up to $\sim$0.15~dex, all of the observed [Mg/Fe] abundance ratios appear to suggest a gradual increase with stellar mass.  Once again, we caution here that this trend is somewhat dependent on the size of the bins used to construct the histogram and the number of galaxies present at low stellar masses.
 The offset between the observed and simulated [Mg/Fe] ratios may point to an over production of Mg relative to Fe which is especially true for the lower mass systems. An increase in the SNIa rate by a factor of three and a half for all galaxies (as explored for star forming MW-like galaxies in Figure~\ref{fig:small}) would bring the normalization of the observed and simulated [Mg/Fe] into broad agreement, but would not address any possible trend at low stellar masses for which we have a limited number of galaxies in our sample.

In addition to uncertainties in the SNIa rate that we explored above, assumptions about the progenitors of SNII can alter the simulated [Mg/Fe] ratios. The assumed lower mass limit for core-collapse supernovae in IllustrisTNG is 8~${\rm M}_\odot$, a generous lower limit for the progenitors of a SNII \citep{smartt2009, ibeling2013, woosley2015, tug2016}. Given the dependence of SNII yields on both progenitor mass and metallicity, a post-processing modification of the SNII rate requires more care than a simple scaling for altering its impact on the [Mg/Fe] ratio in our Milky Ways. In general, one can state that a decrease in SNII would result in overall lower magnesium ratios with respect to iron, as the former is generated predominately in SNII while the latter in SNIa.

Finally, while abundances from detailed spectra of individual stars observed in the Milky Way lend themselves to a more straightforward comparison to simulated abundance ratios based on star particles within a simulation, many effects including measurement errors, accurate atmospheric modeling, and observational completeness can influence the robustness of such a comparison \citep{wyse1998, ramirez2013, nicholls2017}. Thus, for the purposes of this paper, we focus our discussion on the abundance patterns within simulated star forming Milky Way mass galaxies.  However, we caution that while overall trends can be reproduced, a more detailed processing of our simulated data is required to perform a comprehensive statistical comparison between simulated and observed abundance trends.

\section{Characterizing the R-Process and $\alpha$-enrichment in Milky Way-like galaxies} \label{section:mwabundances}
 
The stellar light, maximum [Eu/Fe] distribution, and [Mg/Fe] distribution are displayed in face-on images in the top, middle, and bottom full rows of Figure~\ref{fig:mwimages}, respectively, for a sample of five Milky Way mass galaxies from IllustrisTNG.  Additionally, edge-on views of [Eu/Fe] are shown in the bottom half panels.  The stellar light images are created from the SDSS r-, g-, and i-bands based on the \citet{bruz2003} stellar population synthesis models. Since the Milky Way galaxy selection criteria applied here select only on stellar mass and star formation rate, the resulting Milky Way mass galaxies have a variety of morphologies and sizes, as seen in the top row of Figure~\ref{fig:mwimages}. The saturated yellow regions in the center row of Figure~\ref{fig:mwimages} indicate the sites of high [Eu/Fe] ratios. While some sites correspond to large regions of stars, many are more uniformly distributed throughout the galaxies, agreeing qualitatively with observations of the MW showing highly enhanced europium stars in the halo \citep{cohen2004, venn2004, barklem2005}.

The overall [Mg/Fe]  ratio does not show as large variations in overall abundance as that of [Eu/Fe], and more closely follows recent star formation sites (blue regions in the top row), as depicted by the green maps in the bottom row of Figure~\ref{fig:mwimages}. The relatively uniform distribution of magnesium is a result of the predominant production channel being stellar winds or SNII associated with massive stars.  Massive stars are uniformly distributed with stellar mass throughout our simulated galaxies. The lack of stochasticity (either in time or space) in the production of Mg results in an evenly mixed distribution of Mg within the galaxy. The uniformity of Mg throughout our simulated galaxies stands in stark contrast to the regions of very high [Eu/Fe] generated by the stochastically injected NSNS merger products seen in the middle row panels of Figure~\ref{fig:mwimages}.

To quantitatively address the dispersion in [Eu/Fe] as a function of metallicity, Figure~\ref{fig:eufeWithobs} shows the distribution of [Eu/Fe] vs. [Fe/H] for star particles bound to all Milky Way mass galaxies in IllustrisTNG. There is significantly more dispersion at low metallicities in the [Eu/Fe] ratio as shown in Figure~\ref{fig:eufeWithobs} when compared against the [Mg/Fe] abundance ratio depicted in Figure~\ref{fig:small}. Observational data from Milky Way stars is indicated with red lines while the simulated contour levels are marked with white lines. Additionally, we highlight several observations broken into their halo \citep[small symbols,][]{cohen2004,venn2004,barklem2005}, thin disk \citep[medium symbols,][]{reddy2003,venn2004}, and thick disk \citep[large symbols,][]{reddy2003,venn2004} components. We restrict our comparison to the region $\mbox{[Eu/Fe]} > -0.5$ as observations become increasingly more incomplete below this level \citep[e.g.,][]{jacob2013}.
 
Comparison between the highlighted observations and the two dimensional histogram from IllustrisTNG Milky Way-sized galaxies suggests that there is broad overall agreement between the simulations and observations, particularly with respect to the significantly increased dispersion in the [Eu/Fe] values found for low metallicity stars. Moreover, both the simulations and observations find that stars with the largest [Eu/Fe] enhancement preferentially have lower metallicities.    

We quantify the morphological origin of the large [Eu/Fe] spreads in our Milky Way mass galaxies in Figure~\ref{fig:spatialDist}, which depicts the stellar distribution of [Eu/Fe] vs.~distance from the plane of the galaxy in units of the stellar half mass radius. Here we have defined the plane of the disk to lie perpendicular to the total stellar angular momentum vector. As shown in Figure~\ref{fig:spatialDist}, larger enhancements of [Eu/Fe] in the lowest metallicity bin reside both close to and far from the disk of the galaxy, while those in the highest metallicity bin tend to cluster more towards the center of the galaxies. This suggests a greater existence of both lower metallicity and highly [Eu/Fe] enhanced stars in the haloes than in the thin and thick disk populations. These results are broadly consistent with both the observations plotted in Figure~\ref{fig:eufeWithobs}, which show increasing [Eu/Fe] dispersions for stars living further out in the halo of the Milky Way, and with simulated results from r-process calculations in zoom-in models \citep{vandevoort2015}.

\section{Origin of High [Eu/Fe] at Low Metallicities} \label{section:origins}

The difference in spatial distribution between highly [Eu/Fe] enhanced star particles and those which are more moderately enriched hints at a potential difference in their formation or assembly histories. In this section, we explore the possible scenarios that can explain the origin of highly europium enhanced stars. We specifically consider stellar birth location, galaxy properties at $z=0$, and gas and star formation properties at earlier redshifts ($z=2-4$).

\subsection{Assembly history}

One possibility is that high and low europium enhanced stars were formed in different haloes that later merged. We explore this possibility in Figure~\ref{fig:eufeinex}, which shows the distributions of [Eu/Fe] for three different [Fe/H] ranges. In addition to breaking our simulated [Eu/Fe] distributions into [Fe/H] bins, we generate separate one dimensional histograms for both the ``in-situ'' (cyan lines) and ``ex-situ'' (red lines) populations. Here, ``in-situ'' stars are those formed within the main halo, while ``ex-situ'' stars are those accreted by mergers \citep{rodriguez2016}.

We find that, for high [Fe/H] stellar populations (right panel), the in-situ and ex-situ stellar populations are broadly in agreement with each other and the observations. However, some small discrepancies are notable -- in general, we predict more stars at low [Eu/Fe] than observed at all [Fe/H] bins, and more high [Eu/Fe] stars than observed in the highest [Fe/H] bin.  In addition, while there are many galaxies with a sizable fraction of their mass in highly europium enhanced stars at low [Fe/H], we tend to underpredict the amount of stellar mass in stars with $\mbox{[Eu/Fe]} > 1.0$.  These differences are likely due to a combination of both observational and computational constraints.  

Observations of poorly enhanced [Eu/Fe] stars are difficult, making it likely that observational studies miss many of the stars we predict with $\mbox{[Eu/Fe]} \sim -0.5$ \citep[e.g.,][]{jacob2013}.  In addition, comparisons of numbers of stars with highly enhanced [Eu/Fe] ratios are subject to large uncertainties, both in observations \citep[$\sim$0.2-0.5~dex, e.g.,][]{jacob2015}, and in simulations (see~Section~\ref{section:observations}). Finally, at higher metallicities, some europium lines can become blended with other metal species, making precise determinations of abundances more difficult \citep[e.g.,][]{koch2002}.  Taking these uncertainties in consideration, several trends in the simulations are nonetheless of interest.

At high [Fe/H], the ex-situ stars are slightly offset toward higher [Eu/Fe] with respect to the in-situ stars, while at low [Fe/H], this trend disappears, or can even reverse in individual galaxies (not shown).  In general, the variations from galaxy to galaxy are greater in the ex-situ population than in the in-situ population as shown by the relatively larger error bars for the ex-situ component (red) in Figure~\ref{fig:eufeinex}. The difference in these populations motivates a study of whether the creation of high [Eu/Fe] stars at low [Fe/H] is due to the assembly histories of each MW halo as the differences between in-situ and ex-situ populations seem to suggest.
 
To test the possible relationship between assembly history and europium enhancement, Figure~\ref{fig:fractions} shows the dependency of the fraction of present-day highly europium enhanced stars on galaxy properties. As shown in the left-most panel of Figure~\ref{fig:fractions}, the ratio of ex-situ to in-situ stars is not correlated with the fraction of a MW's stellar mass that is made up of stars with [Eu/Fe] larger than the 75\% contour shown in Figure~\ref{fig:eufeWithobs} ($\mbox{[Eu/Fe]} \gtrsim 1.3$). 
If the assembly history of these individual stars was the driving factor in determining what haloes hosted large [Eu/Fe] enhanced stellar populations, then a trend of increasing (decreasing) ${\rm f[Eu/Fe]_{75\%}}$ with an increasing fraction of ex-situ stars should be present if accreted stars (local stars) are dominating the [Eu/Fe] enhancements. However no such trends are present in the left panel of Figure~\ref{fig:fractions}.  The results presented here do not depend strongly on the exact confidence interval chosen.

In addition, the [Mg/Fe] abundance is not correlated with the ratio of ex-situ to in-situ stars, as shown by the gray contours in the left panel of Figure~\ref{fig:fractions}.  The overall larger vertical spread of the [Eu/Fe] enhancement fractions in comparison to those in [Mg/Fe] points to a greater inhomogeneity of [Eu/Fe] in the stellar populations owing to its stochastically injected production as a function of time. These results are in agreement with those of the zoom-in simulations of \cite{vandevoort2015} which find their low [Fe/H] and [Eu/Fe] stars are also formed in-situ and migrate to larger radii through dynamical effects. In addition, there are no obvious trends of an enhancement fraction with the number of major or minor mergers a halo has experienced during its assembly (not shown).

\subsection{Galaxy properties at $z=0$}

In addition to assembly history, many present-day halo properties show little correlation with the abundance of enhanced europium stars in the IllustrisTNG Milky Ways. Two examples of this are shown in the last two panels of Figure~\ref{fig:fractions}. In general, the fraction of europium enhanced stars is relatively independent of present day halo properties, as is shown in the middle panel of Figure~\ref{fig:fractions}, which depicts the spread in $\rm{f[Eu/Fe]_{75\%}}$ and $\rm{f[Mg/Fe]_{75\%}}$ as a function of halo mass. The right panel of Figure~\ref{fig:fractions} shows the relative independence of the fraction of enhanced stars with the mass-weighted mean stellar age of each Milky Way.

\subsection{Interplay of early Universe gas and galaxy properties}

In the left panel of Figure~\ref{fig:agespread}, we plot the [Eu/Fe] abundance spread with respect to stellar age for every star particle in the 864 Milky Ways in the TNG100 simulation. The typical age of the europium enhanced stars in our simulation of 10-12~Gyrs is broadly consistent with estimates obtained from Milky Way zoom simulations \citep{shen2015, vandevoort2015}, and overlaps somewhat with both the peak in the cosmic star formation rate (left panel of Figure~\ref{fig:cosmicRatesmwRates}) and peaks in the star formation rates of individual Milky Ways (right panel, Figure~\ref{fig:cosmicRatesmwRates}).

Over this time range, the distribution of cosmic gas abundances changes. The gas phase [Eu/Fe] abundance becomes larger and increasingly homogeneous with time as the gas enriched with the stochastically distributed NSNS mergers mixes with less enriched gas, as shown by the more strongly peaked [Eu/Fe] distribution at $z=0$ than at $z=4$ in the right panel of Figure~\ref{fig:agespread} (dark lines).  This implies  early times are more favorable for the creation of europium enhanced stars.  This is in contrast to the evolution of the gas phase [Mg/Fe] ratio (light lines in Figure~\ref{fig:agespread}, right panel) which decreases and grows slightly less homogeneous with time.  This is likely due to the increase of enrichment by Mg and Fe from stellar winds and SNIa at later times.

\begin{figure}
\centering
\includegraphics[width=0.45\textwidth]{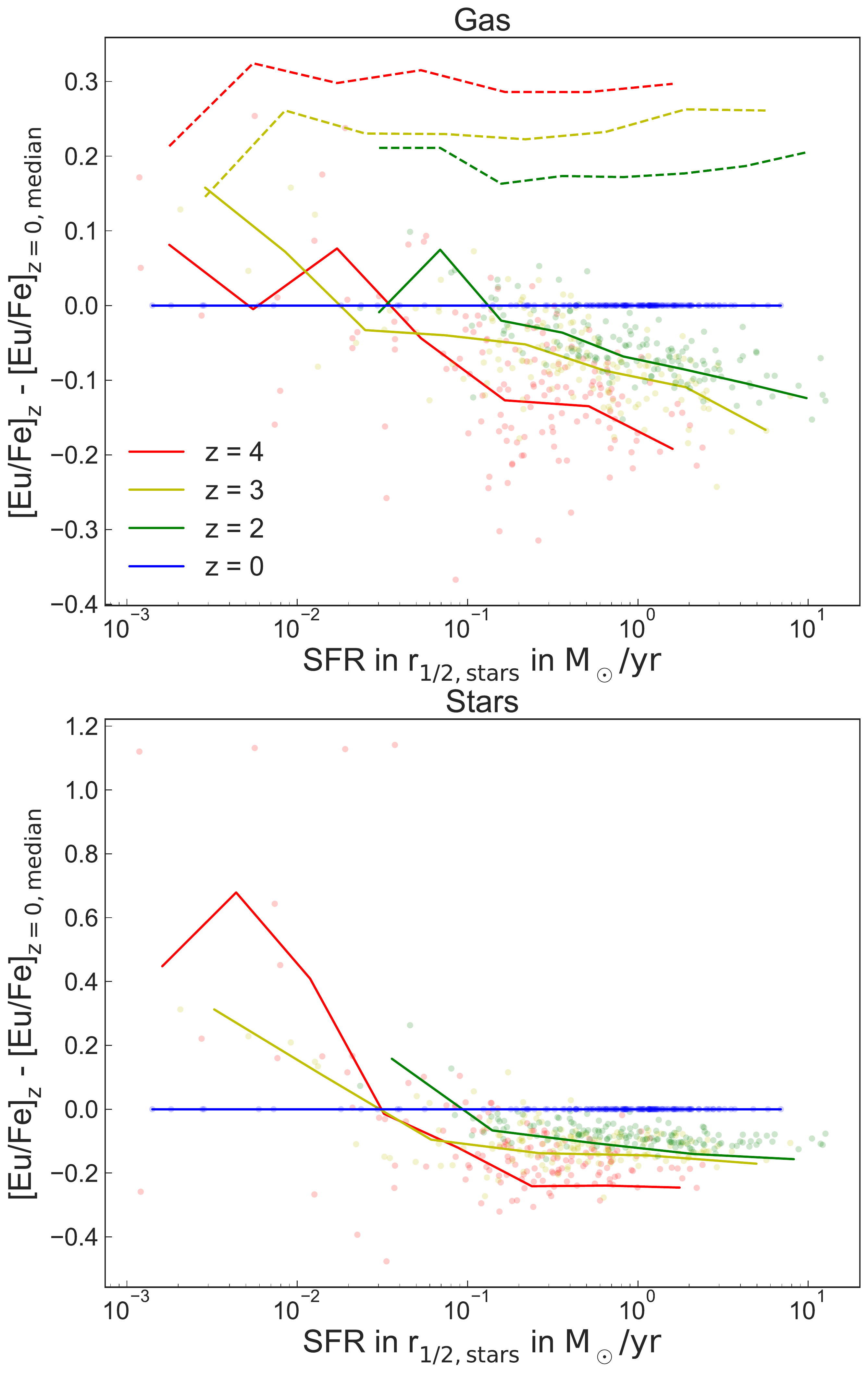}
\caption{Changes in gas phase [Eu/Fe] with redshift and SFR, and its translation to stellar phase [Eu/Fe] at $z=0$ in MW-sized galaxies.  The top panel shows the gas mass-weighted [Eu/Fe] in a galaxy at several different redshifts -- lower SFR at the higher redshifts leads to higher [Eu/Fe] and the trend flattens out as redshift decreases.  
For comparison, the trends for gas phase [Mg/Fe] are depicted with dashed lines showing no such trend with redshift or SFR.
The bottom panel shows how the stellar phase [Eu/Fe] tracks with SFR at different redshifts.  As more stars form at lower redshifts, the trend of decreasing [Eu/Fe] with increasing SFR is washed out as the gas phase europium abundance homogenizes.  Each point is the mass-weighted average [Eu/Fe] in the gas or stellar phase of a galaxy with its $z=0$ value subtracted.  Thus the dark blue lines show the normalized $z=0$ values of [Eu/Fe] in gas and stellar phases which are zero by construction.}
\label{fig:eufesfr}
\end{figure}

The relation between high europium stellar contamination and early star formation seems in contradiction with the results shown in the last panel of Figure~\ref{fig:fractions} which shows little correlation between the mass-weighted stellar age of a halo and the fraction of europium enhanced stars it hosts. This apparent contradiction points to the interplay between gas inhomogeneity and star formation in determining the levels of stellar europium enhancement.  Figure~\ref{fig:eufesfr} shows this interplay in both the gas phase [Eu/Fe] and the fraction of europium enhanced stars. In the top panel, each point represents the gas mass-weighted average for [Eu/Fe] in each of the 864 MW-sized haloes at various redshifts.  There is a clear trend of increasing [Eu/Fe] enhancement with decreasing star formation rates.  This is in contrast to the [Mg/Fe] distribution, which shows no such correlation (dashed lines in the top panel of Figure~\ref{fig:eufesfr}) and simply decreases in overall normalization as time progresses and more and more gas is polluted with iron-rich SNIa ejecta. The trend of decreasing [Eu/Fe] with SFR is replicated in the mass-weighted stellar [Eu/Fe] distribution as well, as shown in the bottom panel of Figure~\ref{fig:eufesfr}.

The impact of the SFR on decreasing the [Eu/Fe] abundance ratio is likely two fold. More star formation results in more enrichment from supernovae, which induces more mixing in the gas at $z \approx 2-4$, when the high [Eu/Fe] stars would otherwise be forming from inhomogeneous gas, as pointed out in Figure~\ref{fig:agespread}. Moreover, more star formation increases the enrichment by SNII and SNIa, which in turn increases the iron abundance, thereby decreasing the local [Eu/Fe] ratio of the gas. 
Because the overall [Eu/Fe] ratio increases with time, as shown in the right panel of Figure~\ref{fig:agespread}, it is likely the first effect dominates.

\section{Resolution Effects} \label{section:discussion}

While resolution effects in magnesium and iron mostly originate from the different amount of stars formed, and thus the number of SNII and SNIa explosions, the interplay between resolution and our sub-grid europium distribution prescriptions results in  greater resolution variations for the [Eu/Fe] than [Mg/Fe] ratio. However, the strongest effect is restricted to the overall normalization, while the dispersion of [Eu/Fe] at low [Fe/H] with respect to high [Fe/H] remains stable.

A first effect on the europium distribution in our simulated Milky Ways comes from our assumptions for the size of the cooling mass. One can approximate this cooling mass based on the final momentum of the NSNS merger material during its momentum conserving phase, resulting in a dependence on the local ISM density ${n_{\rm ISM}}$, ejecta energy $E$, local sound speed ${v_{\rm ISM}}$, or local turbulent velocity  $\sigma \propto n_{\rm ISM}^{-1/7} E^{13/14} {v_{\rm ISM}}$ \citep{cioffi1988, vandevoort2015, montes2016}. As noted in \cite{piran2014} and \cite{vandevoort2015}, while typical values for the ISM lead to a cooling mass of $M_{\rm mix} \approx 10^{3.5}-10^{4} \, {\rm M}_\odot$ \citep{montes2016}, assumptions about the local conditions of the gas, in particular with respect to ${v_{\rm ISM}}$, can vary greatly in the presence of intense turbulent mixing from star formation, galactic wind feedback, and disk rotation instabilities. 

\begin{figure*}
\centering
\includegraphics[width=1.0\textwidth]{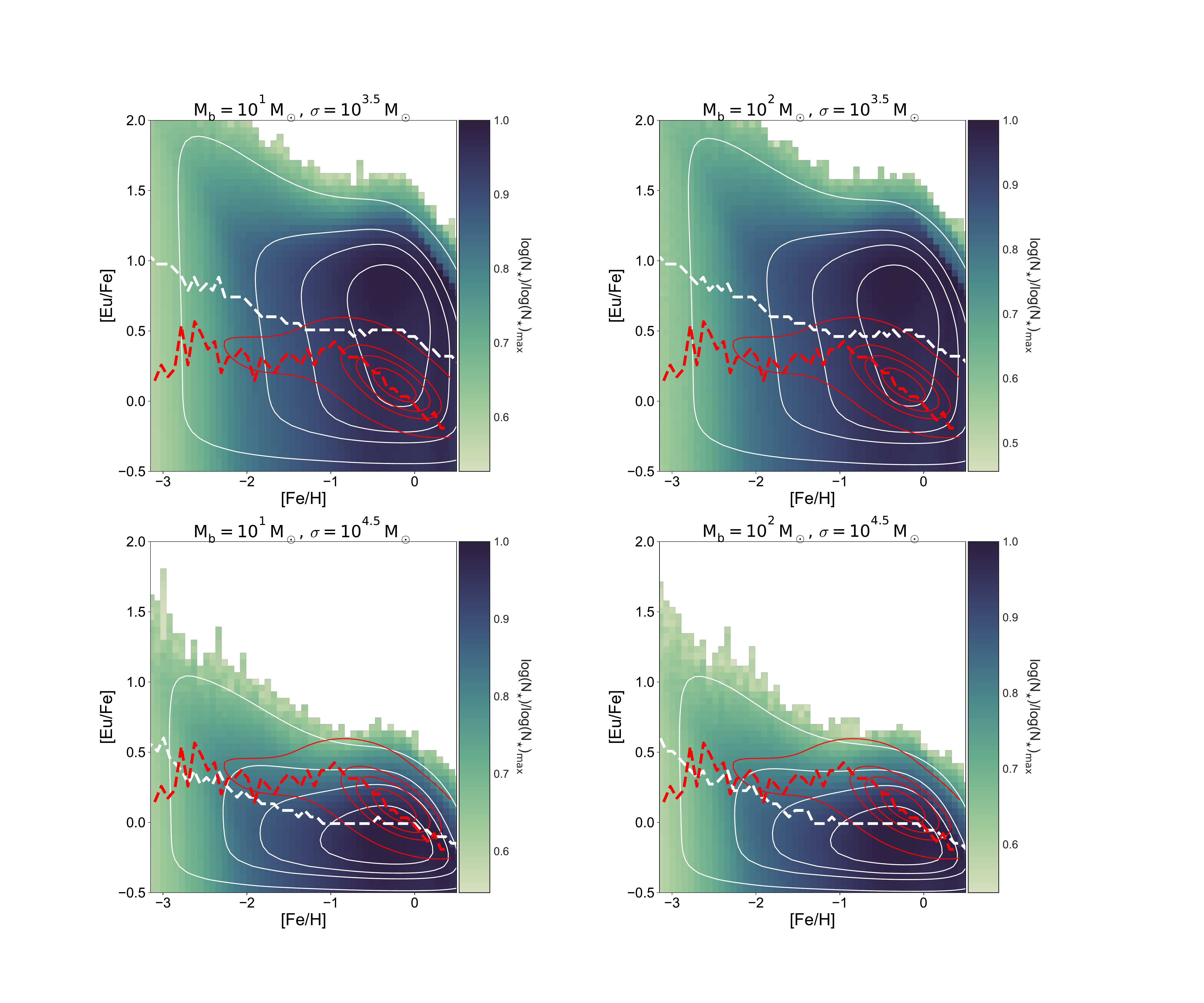}
\caption{Changes in the [Eu/Fe] distribution for order of magnitude changes in both the size of our sub-mass bins, $M_b$, and the cooling mass spread parameter, $\sigma$.   White (red) contours are 10\%, 25\%, 40\% and 75\% contour levels of the simulated (observed) data based on a kernel-density estimate while dashed lines show the medians of the distributions in each [Fe/H] bin.  As $\sigma$ is changed from $10^{3.5} \, {\rm M}_\odot$ (top row) to $10^{4.5} \, {\rm M}_\odot$ (bottom row), the overall normalization of the distribution shifts vertically by approximately $\lesssim 1.0$~dex.  A change in the sub-mass bin size results in negligible variation of the fiducial distribution. }
\label{fig:eufetests}
\end{figure*}

In addition to changes due to subgrid turbulent mixing, the metallicity of the surrounding gas can affect the cooling efficiency of the remnant as well, decreasing the value of the cooling mass at higher metallicities, $M_{\rm{cool}} \propto Z^{-3/7}$ \citep{cioffi1988, thornton1998, martizzi2015, macias2016}. While our Gaussian prescription described in equation~(\ref{eq:gaussian}) allows for mixing outside the cooling mass, instead of a strict cut-off at $M_{\rm mix} = 10^4 \, {\rm M}_\odot$, the assumed dispersion, $\sigma$, of our Gaussian distribution can influence the shape of the europium enhancement in our Milky Way sized galaxies. This result, shown by comparing the top and bottom panels in Figure~\ref{fig:eufetests}, manifests itself predominately as a change in the overall normalization by $\lesssim$1.0 dex if the cooling mass changes by an order of magnitude.  We also investigated the effects of decreasing the size of our sub-mass bins from $10^2 \, {\rm M}_\odot$ (right panels) to $10 \, {\rm M}_\odot$ (left panels), which turns out to be negligible. 

\begin{figure*}
\centering
\includegraphics[width=1.0\textwidth]{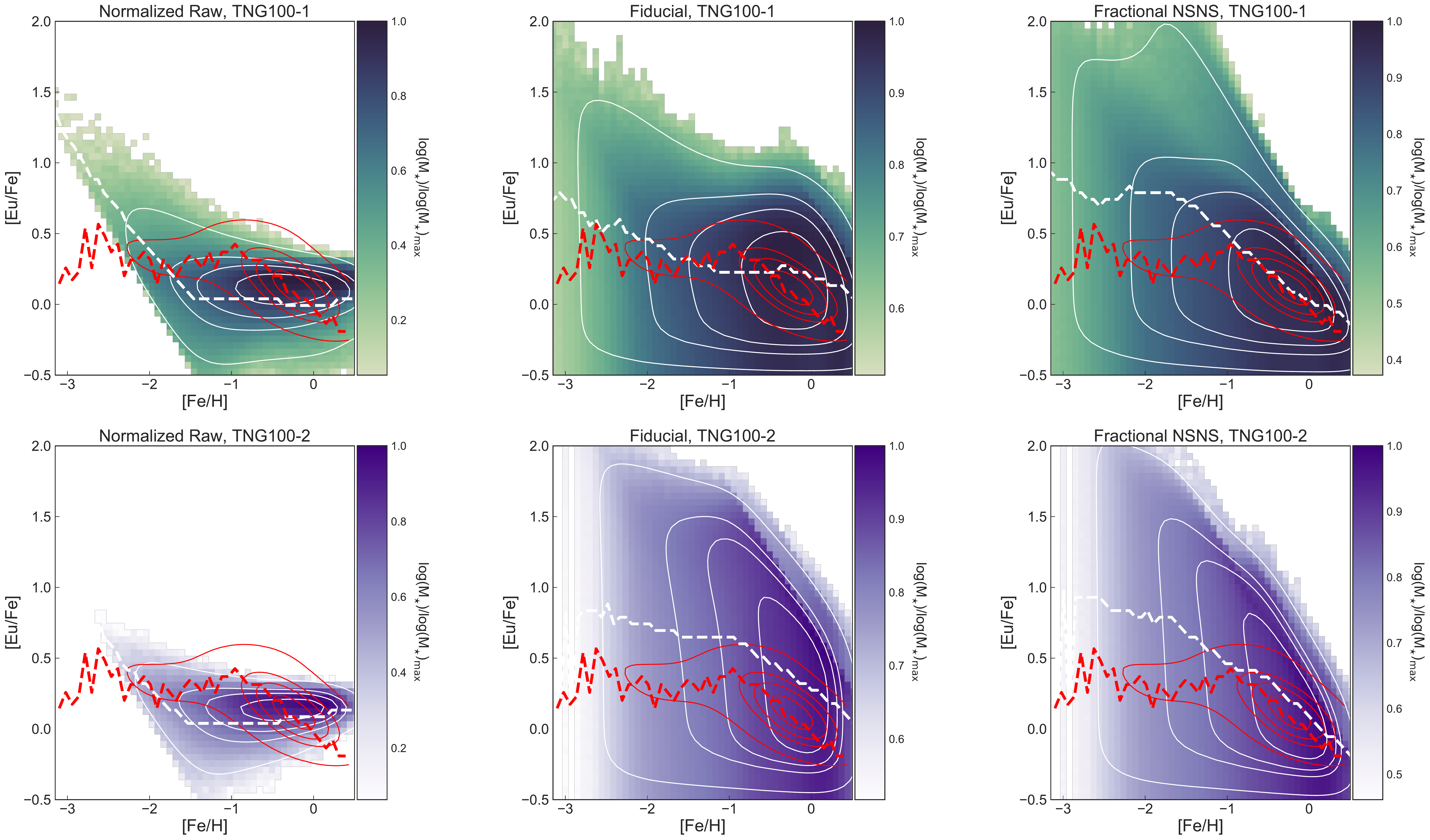}
\caption{The [Eu/Fe] distribution in the TNG100 box is influenced by our choices of post processing cooling mass distribution parameters.  The left column depicts the un-postprocessed distribution.  White (red) contours are 10\%, 25\%, 40\% and 75\% contour levels of the simulated (observed) data based on a kernel-density estimate while dashed lines show the medians of the distributions in each [Fe/H] bin.  Here, the simulated points are shifted up by 1.3~dex to better fit the observed data for the purposes of plotting.  At the highest resolution, the spread in [Eu/Fe] at low metallicities with respect to that at $\mbox{[Fe/H]} \sim 0$ is nearly reproduced.  However the shape at $\mbox{[Fe/H]}\sim 0$ is far too flat, and the effects of our mass fraction cut-off of $2.5 \times 10^{-10}$ are clearly visible as a line in the distribution at low [Fe/H]. The overall shape of the distribution is better reproduced by our fiducial post-processing model at high resolutions in the upper middle panel, however, this model over-produces Eu as resolution decreases with the fiducial choice of cooling mass parameter of $\sigma = 10^4 \, {\rm M}_\odot$.  The ``fractional NSNS'' post-processing model shown in the right most column provides better consistency across resolutions, however, it overproduces Eu with $\sigma = 10^4 \, {\rm M}_\odot$.}
\label{fig:eufeConvergence}
\end{figure*}

Besides the assumptions about the size of the cooling mass, the treatment of fractional amounts of NSNS merger material in star forming gas can have an effect on the convergence of our sub-grid prescription. Without any post processing, the spread of [Eu/Fe] at low [Fe/H] with respect to that at $\mbox{[Fe/H]} \sim 0$ nearly fits the observed data in our highest resolution simulation after a shift of approximately 1.3~dex vertically, as shown by the left column of Figure~\ref{fig:eufeConvergence}.  While the dispersion is less evident in lower resolution simulations, at our highest resolution it is prominent even though the cooling mass is two or three orders of magnitude smaller than the gas cell resolution. Given the uncertainties in the rate of NSNS mergers, the cooling mass, the amount of r-process material ejected, and the percentage of ejecta in europium, such a shift, while large, is not ruled out by current observational constraints. Thus, we emphasize here that any changes in our prescription affect predominately the (uncertain) normalization of our distribution, while the overall relations between dispersions at low and high [Fe/H] are hardly influenced thus making the NSNS merger r-process progenitors a viable explanation of the europium spreads observed in the Milky Way. 
Additionally, we performed our analysis on trends of high [Eu/Fe] enhancement with various stellar and galactic properties (encapsulated in Figures \ref{fig:spatialDist} - \ref{fig:eufesfr}) without a subgrid formalism and the results are consistent with those which include the subgrid prescription.  Thus, any subgrid treatment affects only the size and normalization of the [Eu/Fe] spread at low [Fe/H] and not how the spread depends on a galaxy's formation history.

In our fiducial model, we assume that any material from a partial NSNS merger within a star particle is distributed following the same Gaussian given in equation~(\ref{eq:gaussian}), albeit with a smaller normalization equal to the fraction of NSNS ejecta, as depicted by the smaller/fainter Gaussian bump in Figure~\ref{fig:subgrid}. As the resolution of our simulation increases, this results in more cells being populated with fractional merger material and less cells with whole NSNS mergers, and thus our fiducial prescription begins to underpredict the [Eu/Fe] in more highly resolved simulations. This can be clearly seen in the middle column of Figure~\ref{fig:eufeConvergence} where our fiducial model tends to overpredict the [Eu/Fe] ratio in lower resolution boxes. Instead of assuming that fractional NSNS mergers are distributed following the same overall Gaussian of equation~(\ref{eq:gaussian}) with varying normalizations, one can assume that all NSNS mergers share the same normalization and fractional mergers are simply fractions of one Gaussian spread across many cells. This results in a more consistent [Eu/Fe] distribution across all resolutions, as shown by the right columns of Figure~\ref{fig:eufeConvergence}, however, at an overall higher normalization than in our fiducial model. This prescription serves as an upper limit on the amount of r-process mass that may be distributed in a sub-mass bin for a given value of $\sigma$ in equation~(\ref{eq:gaussian}).  Both, larger values of the cooling mass and a ``fractional" distribution of NSNS merger mass, do not strongly affect any of the conclusions presented here.
 
\section{Conclusions} \label{section:conclusions}

We have presented the first results of the new chemical evolution implementation in the IllustrisTNG suite of simulations, the successor of the Illustris project with an updated galaxy formation model. The updated model includes an improved advection scheme for the 9 individually tracked metals, revised yield tables, metal production site tagging, and explicit tracking of ejecta from neutron star -- neutron star mergers. It also includes important improvements in the treatment of feedback processes from star formation and supermassive black hole accretion, yielding substantial progress in the predicted galaxy properties.  

We began our analysis by comparing the simulated $\alpha$-abundances with those observed in a variety of galaxies -- both from quiescent stacked SDSS spectra and individual stars in the Milky Way -- and emphasized some of the recent issues relevant to comparing observational and simulated abundance ratios. We then focused on neutron star -- neutron star driven europium enhancements within the Milky Way in our cosmological context, paying particular attention on how this stochastically injected process serves as a proxy for the efficiency of gas mixing during periods of intense star formation. Here, comparisons between observed and simulated Mg abundances are used as indicators of the effectiveness of our model on large and small scales, while comparisons between simulated and observed europium abundances in the Milky Way highlight the effects of mixing in our simulations -- a key component in any theory of cosmological chemical evolution.
 
\noindent Our main findings are:
\begin{itemize}

\item A robust comparison between observed and simulated abundances of magnesium and iron is non-trivial, and we thus caution that assumptions about the processing of simulated abundance ratios into mock observations can affect the comparison. We further discuss how the assumed rates of SNIa and SNII can affect abundance ratios in Milky Ways.  We emphasize that while broad trends in abundance ratios can be recovered, a strict comparison with observations requires a more careful treatment than that discussed here.

\item From a set of Milky Way sized galaxies we reproduce the shape and spread of the observed [Eu/Fe] and [Mg/Fe] abundances in the Milky Way. The large [Eu/Fe]  enhancements at low [Fe/H] are reproduced with NSNS mergers as the sole enrichment process, with some deviations from the Milky Way observations at higher [Fe/H], dependent upon choices for the distribution of NSNS merger material in post processing.  The Milky Way [Mg/Fe] ratio depends both on methods for comparing between observations and simulations, and assumptions about magnesium yields in SNII and SNIa rates.  While our [Mg/Fe] ratio overlaps with the observed data, a better fit is recovered if the rate of SNIa in our Milky Ways is increased by a factor of three and a half.

\item Highly europium enhanced stars, those with $\mbox{[Eu/Fe]}\gtrsim 1.3$, originate from older periods of star formation than those without strong enhancements and are predominately formed for redshifts $z\sim 2-4$, during a period of time when the cosmic gas phase distribution of [Eu/Fe] was less homogeneous than it is presently, allowing for stars to form from highly europium enhanced pockets of gas.  In contrast to the increase of the gas phase homogeneity of [Eu/Fe], the [Mg/Fe] gas phase ratio becomes less homogeneous with time, likely owing to the increase in Mg and Fe from sources other than SNII (stellar winds and SNIa).

\item We find the fraction of [Eu/Fe] and [Mg/Fe] enhanced stars in our simulated galaxies is robust to changes in total halo mass, mass-weighted population age, and formation history.

\item Gas phase europium enhancements are instead anti-correlated with the star formation rate of their host galaxy at early times, suggesting that vigorous star formation homogenizes regions of europium enhanced gas, limiting the regions that can turn into highly europium enhanced stars. This scenario is born out in trends of simulated gas and stellar phase [Eu/Fe] ratios.  There is no such correlation of [Mg/Fe] with star formation rate and redshift.

\item Finally, while the large spread of [Eu/Fe] at low [Fe/H] with respect to that at high [Fe/H] is robust to changes in both resolution and the post-processing subgrid model for the spread of NSNS merger material, the overall normalization of the [Eu/Fe] ratio depends on both these factors.  This is in contrast to changes in [Mg/Fe] with resolution, which are relatively small.  This reflects upon both the different channels of enrichment -- one stochastic (Eu), the other more homogeneous (Mg) -- and the necessity of a subgrid modeling for NSNS mergers.

\end{itemize}

Future IllustrisTNG simulations and follow up zoom-in calculations will begin to resolve the cooling mass scale of neutron star -- neutron star mergers, providing a robust test of the results presented in this paper.  In addition, the wealth of chemical evolutionary information available in the new IllustrisTNG simulation suite will allow us to probe both galactic- and cosmological-scale mass and metal return more fully in upcoming works.

\section*{Acknowledgements}

The authors thank Vicente Rodriguez-Gomez for the use of his galaxy catalogs, Maria Bergemann for useful conversations and the referee, Dr.\ E.\ Scannapieco, for his useful comments.  Simulations were run on the HazelHen supercomputer at the High-Performance Computing Center Stuttgart (HLRS) as part of project GCS-ILLU of the Gauss Centre for Supercomputing (GCS). VS, RP, and RW acknowledge support through the European Research Council under ERCStG grant EXAGAL-308037.  VS also acknowledges support through subproject EXAMAG of the Priority Programme 1648 Software for Exascale Computing of the German Science Foundation. SG and PT acknowledge support provided by NASA through Hubble Fellowship grant HST-HF2-51341.001-A and HST-HF2-51384.001-A, respectively, awarded by the STScI, which is operated by the Association of Universities for Research in Astronomy, Inc., for NASA, under contract NAS5-26555. SG also acknowledges the support of the Flatiron Institute supported by the Simons Foundation. LH acknowledges support from NASA grant NNX12AC67G and NSF grant AST-1312095. ERR acknowledges the support of the Packard Foundation and UCMEXUS (CN-12-578) and MV acknowledges support through an MIT RSC award, the support of the Alfred P. Sloan Foundation, and support by NASA ATP grant NNX17AG29G. JPN acknowledges support of NSF AARF award AST-1402480.

\bsp	





\bibliographystyle{mnras}

\bibliography{bib_rprocess}


\label{lastpage}
\end{document}